\documentclass[sigchi]{acmart}

\usepackage{xcolor}

\newcommand{\change}[1]{{\textcolor{black}{#1}}}

\usepackage{xurl}
\usepackage{xspace}
\newcommand{\system}{AltCanvas\xspace} 
\newcommand{\systems}{AltCanvas's\xspace}

\usepackage{mdframed}
\usepackage{setspace}
\usepackage{pdfcomment}
\usepackage{balance}
\definecolor{lightbrown}{RGB}{255,249,241}
\newenvironment{promptbox}
    {\begin{mdframed}[backgroundcolor=lightbrown,linecolor=white]
        \begin{small}
            \begin{spacing}{0.9}}
    {\end{spacing}
    \end{small}
    \end{mdframed}
    }

\usepackage{array}
\newcolumntype{L}{>{\centering\arraybackslash}m{3cm}}

\AtBeginDocument{%
  \providecommand\BibTeX{{%
    \normalfont B\kern-0.5em{\scshape i\kern-0.25em b}\kern-0.8em\TeX}}}

\copyrightyear{2024} 
\acmYear{2024} 
\setcopyright{acmlicensed}\acmConference[ASSETS '24]{The 26th International
ACM SIGACCESS Conference on Computers and Accessibility}{October 27--30,
2024}{St. John's, NL, Canada}
\acmBooktitle{The 26th International ACM SIGACCESS Conference on
Computers and Accessibility (ASSETS '24), October 27--30, 2024, St. John's,
NL, Canada}
\acmDOI{10.1145/3663548.3675600}
\acmISBN{979-8-4007-0677-6/24/10}

\begin{document}

\title[\system: Visual Content Creation with Generative AI for BVI People]{\system: A Tile-Based Image Editor with Generative AI for Blind or \change{Visually} Impaired People}



\author{Seonghee Lee}
 \affiliation{%
  \institution{Stanford University}
  \country{USA}
 }
 \email{shlee@cs.stanford.edu}

 \author{Maho Kohga}
 \affiliation{%
  \institution{Stanford University}
  \country{USA}
 }
 \email{mkohga@stanford.edu}

 \author{Steve Landau}
 \affiliation{%
  \institution{Touch Graphics}
  \country{USA}
 }
 \email{sl@touchgraphics.com}

 \author{Sile O'Modhrain}
 \affiliation{%
  \institution{University of Michigan}
  \country{USA}
 }
 \email{sileo@umich.edu}

\author{Hari Subramonyam}
 \affiliation{%
  \institution{Stanford University}
  \country{USA}
 }
 \email{harihars@stanford.edu}

\renewcommand{\shortauthors}{Lee, et al.}

\begin{abstract}
People with visual impairments often struggle to create content that relies heavily on visual elements, particularly when conveying spatial and structural information. Existing accessible drawing tools, which construct images line by line, are suitable for simple tasks like math but not for more expressive artwork. On the other hand, emerging generative AI-based text-to-image tools can produce expressive illustrations from descriptions in natural language, but they lack precise control over image composition and properties. To address this gap, our work integrates generative AI with a constructive approach that provides users with enhanced control and editing capabilities. Our system, \system, features a tile-based interface enabling users to construct visual scenes incrementally, with each tile representing an object within the scene.  Users can add, edit, move, and arrange objects while receiving speech and audio feedback. Once completed, the scene can be rendered as a color illustration or as a vector for tactile graphic generation. Involving 14 blind or low-vision users in design and evaluation, we found that participants effectively used the \systems workflow to create illustrations.
\end{abstract}

\begin{teaserfigure}
  \centering
   \includegraphics[width=\textwidth]{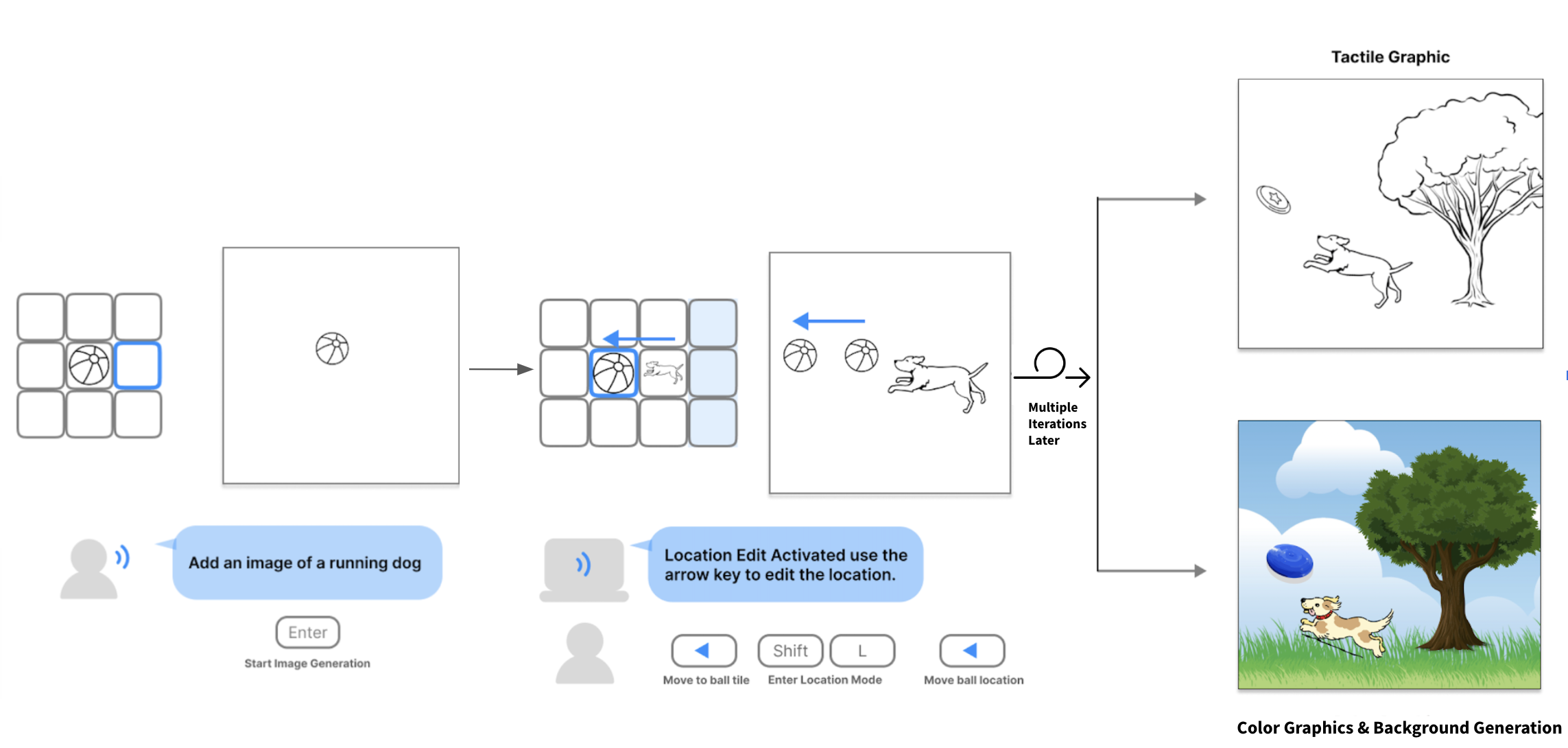}
 \caption{\textbf{ AltCanvas image authoring steps: (A) Image Generation with AltCanvas using a tile-based interface. (B) Image Editing Process in AltCanvas, incorporating sonification and verbal feedback. (C) Final Rendering Process of AltCanvas, producing tactile graphics or enhanced renderings for general audiences.}}
  \label{fig:teaser}
  \Description[]{A diagram of AltCanvas' three steps image authoring steps. A on the left: The tile-based interface can add images by talking to them. B in the middle: Images can be edited through sonification, verbal feedback, and arrow keys. C on the right: The rendering process of black-and-white and color graphics. Step A: It shows a tile-based interface with a highlighted square, suggesting the selection of a tile, and an instruction bubble that reads, "Add an image of a running dog," prompting the start of the image generation. Step B: The interface now includes a line drawing of a running dog next to a ball, with a directional arrow indicating the ability to move the ball. Below, there's a legend explaining the editing controls, such as using arrow keys to adjust the location of objects. Step C: This step is divided into two parts. The upper part shows the final black-and-white tactile graphic, a simple and clear line drawing of a dog chasing a flying disc with a tree in the background. The lower part displays the color graphic with final rendering, where the same scene is depicted in full color with a blue sky, a lush green tree, and the dog joyfully leaping towards the blue flying disc.}
\end{teaserfigure}
\maketitle

\section{Introduction}

Spatial constructs play a crucial role in human cognition and communication~\cite{laurini1992fundamentals}. We rely on spatial information for describing scenes, learning complex concepts in science and mathematics, and artistically expressing ourselves. For instance, when describing a pet-friendly home in a blog article, we can sketch out where to place pet beds, food stations, and play areas showing dogs and cats at play. For sighted users, several tools exist for creating visual graphics across different fidelity, ranging from digital pen-based drawing applications (e.g., SketchPad~\cite{sutherland1963sketchpad}) to direct manipulation tools such as Adobe Illustrator~\cite{AdobeIllustrator} for authoring rich vector graphics. However, based on a recent survey~\cite{zhang2023understanding}, while blind or vision impaired (BVI) people are motivated to create visual content for school, work, and art, current access barriers in content creation tools hinder their ability to do so. Although linguistic references such as ``next to'' or ``jagged'' are commonly used for spatial and structural descriptions, they can also be abstract and ambiguous. The act of \textit{visualizing} allows us to be more precise in articulating spatial information.  

Existing drawing tools designed for BVI users take a guided constructive approach (i.e., line-by-line drawing) and primarily concentrate on basic graphics, such as shapes and geometric figures~\cite{bornschein2017digital}. While this allows a high degree of control within specific drawing tasks, it can be tedious to create more expressive open-ended graphics, for instance, the illustration of the pet-friendly home in the earlier example. On the other hand, tools based on emergent text-to-image generative models~\cite{dhariwal2021diffusion} allow for rich, expressive illustration and digital artwork using text prompts~\cite{chang2023prompt}. However, the trade-off is in the \textit{degree of control}. With a generative AI  approach, users may have to describe entire scenes in natural language and have less control over specific spatial and visual attributes such as composition, size, color, relative distance, orientation, etc. In other words, it would be difficult to generate a highly specific illustration of the intended pet-friendly home. Our own exploration has shown that the models don't always accurately interpret text prompts (e.g., a prompt for \textit{``image with five apples and one apple sliced in half''} resulted in one apple sliced into five pieces). In fact, recent work in generative AI-based image tools for BVI users has looked at question-answering mechanisms to validate the generated output across complex scenes~\cite{huh2023genassist}. Alternatively, creating accessible visual content like printed tactile graphics is also challenging, as direct conversion is often impossible, making iterative design difficult.

Ideally, we would like tools that can leverage generative AI capabilities to create expressive content but also support a constructive approach for better control and composition. In this regard, for sighted users, researchers have explored techniques such as users providing a rough doodle of scene composition to inform the semantics and style of generation~\cite{park2019semantic}. Turning our thoughts into tangible visuals requires such forms of control through iterative representation, feedback, and adjustments~\cite{millar1975visual}. Thus, our motivating question for this work is \textit{``How might we support a constructive approach to authoring visual content while also leveraging the generative capabilities of AI models?''} By conducting a formative study with five blind participants who have created visual content in the past, we learned about current content authoring workflows, associated challenges, and opinions about generative text-to-image models. Specifically, experienced users expressed (1) a need for precise control and guidance to optimize the design, (2) the ability to use pre-existing graphics with options for deleting backgrounds and tools for rescaling and editing, and (3) enhanced usability of generative AI features through verbal descriptions and explanations of graphics without needing to print each iteration. Based on these requirements, we developed \textit{\system}, using an iterative design and evaluation approach. 

\system is a generative AI-powered illustration tool that implements a \textit{dynamic tile-based interface} and sonification features to support the authoring of visual graphics. It consists of a side-by-side layout of a tile view (i.e., an alternative to a direct manipulation drawing canvas --- \system) and an image view to render the image under construction. The tile view is designed based on an understanding of blind spatial cognition and relational aspects of objects in images (i.e., \textit{``the vast majority of visual information is really spatial information~\cite{giudice2018navigating}
''}). At the start of creating a new illustration, the tile view consists of a single tile, and dynamically expands along eight directions to allow adding additional objects to relative locations. As shown in Figure~\ref{fig:teaser}, the user starts by adding a ball object to the canvas using speech-to-text and text-to-image. They then add a dog to the right of it by navigating to the adjacent tile on the right. \system implements speech-based descriptions of the generated image and, through keyboard navigation and sonification, allows creators to compose complex images through resizing, repositioning, and other edit features. Finally, once the image is composed, \system allows creators to generate a representation suitable for printing as tactile graphics or can render full-color graphics for sharing with sighted people. In developing \system, we prototyped a range of interaction techniques for different authoring tasks and conducted a preliminary study to understand the preferred interactions of BVI content creators. Based on feedback, we revised our designs and evaluated the final prototype through a user study. Across the formative and iterative evaluation studies, we worked with 14 participants.

Our main contributions include (1) a tile-based interaction paradigm that provides an alternative representation of the visual canvas, (2) a novel constructive and generative image authoring workflow using speech and sonification, and (3) results from authoring content with \system and tactile graphic evaluation from BVI users.

\section{Related Work}

\subsection{Image Editing Challenges for BVI users}
When editing images, users must continually adjust their input based on visual feedback. However, many current editing software lack the capability to offer real-time visual feedback through screen-readers~\cite{Acosta2018, landay2000, Schaadhardt2021}. This is a serious limitation as visual feedback is crucial for editing operations such as alignment and overlap~\cite{Millar1975VisualEO, Peng2021}. Additionally, complex image editors like Adobe Photoshop, with its 71 keyboard commands, compound the challenge, especially when custom menus or dialog boxes aren't screen reader-friendly~\cite{AdobePhotoshop, Acosta2018, pandey2020, li2021}. Further, image editing tools should be equipped with ``grammar tools'' that assist visually impaired users in navigating the spatial layout and understanding the interactive drawing space, enhancing their grasp of concepts or subject matter~\cite{icalt2016}. These tools should incorporate elements that can be adjusted or combined over time, enabling users to locate drawn objects, identify critical control points, and make modifications or additions to these points as necessary~\cite{landay2000}.

\subsection{BVI Image Editing Interfaces}
Grid-based interfaces can help BVI users make more precise point selections \cite{landay1999}. Prior work has explored variations such as IC2D's grid-based system \cite{landay1999} and haptic display with a grid system that divides the drawing surface into $m \times n$ sections \cite{headley2010}. To navigate around different regions on the grid, keyboard commands coupled with verbalizations of grid locations were used. Alternately, the use of a table structure to convey information has also been used in prior work to make data flow diagrams accessible for BVI students \cite{Sauter2015TeachingTM}. Considerations for table-based interfaces include screen-reader accessibility, but it is challenging to operationalize well even if tables have been properly marked up \cite{Williams2019, Gardiner2016}. Innovative navigational aids like those in TextSL offer BVI users the ability to move through virtual spaces safely and efficiently using natural language for collision-free navigation and relative location information, replacing traditional coordinate systems with directional cues such as ``north'' or ``northeast'' \cite{folmer2009}. This suite of tools and methodologies reflects a growing commitment to enhancing the usability of image editing software for BVI users through multi-sensory and accessible interfaces. Yet there is more to be explored in terms of effective guidelines for creating BVI-accessible editing software.

Image description and tactile feedback are crucial components of accessible image editors for visually impaired users. Early innovations in this field include TDraw, which offered tactile feedback without erasure capabilities~\cite{kurze1996}, and Watanabe's system, which introduced full erasure functionality via a Braille display and stylus~\cite{watanabe2002}. Recent advancements have expanded these capabilities: the Draw and Drag Touchscreen provides interactive manipulation with text-to-speech output for geometric shapes~\cite{Grussenmeyer2015}, ``Playing with Geometry'' utilizes vibrotactile feedback for freehand drawing ~\cite{buzzi2015}, and AudioDraw employs audio feedback to assist in creating educational diagrams ~\cite{Grussenmeyer2016}. Parallel developments in image description systems have further enhanced accessibility. The `RegionSpeak' system enables users to crowdsource object labels and navigate spatial layouts~\cite{zhong2015}, while `TactileMaps' introduces a novel tactile exploration method using a raised-line map overlaid on a multi-touch screen with auditory feedback ~\cite{brock2015}. Leveraging advanced technology, `Image Explorer' employs deep learning to create a multi-layered, touch-based exploration system~\cite{lee2021}. These innovative tools, spanning from tactile interfaces to sophisticated image description systems, significantly improve visually impaired users' ability to interact with and understand visual content through a combination of touch and auditory feedback.

Previous research has highlighted the efficacy of data sonification for rapid data analysis. Notable examples include the sonification of the 2011 Tohoku Earthquake in Japan, various astronomical objects by NASA, and solar winds \cite{arcand2024universe, candey2006xsonify, sondergaard2017sonification}. MathGraphs developed a series of sonification of math graphs \cite{ohshiro2021}. For BVI users who are more sensitive and trained to recognize the sound, the use of sonification provides an intuitive way for users to interact beyond immediate visualization \cite{Pinto2011}. Sonification has also been applied to enhance object recognition within interactive experiences. For example, EdgeSonic employs sonification techniques to emphasize image edge features and distance-to-edge maps \cite{yoshida2011}. Similarly, Melodie has enriched the crafting process for weavers by incorporating sonification \cite{rodriguez2021}. More recently, Symphony introduced an audio-tactile system specifically designed to aid blind weavers. This system uses sounds to denote warping and pitch variations, thus facilitating the creation and perception of textile patterns \cite{das2023}. Extending beyond these applications, sonification has found a place in digital gaming for the blind. ``Hero's Call'' utilizes arrow keys for movement and incorporates a sound radar to help players navigate their surroundings \cite{heros_call}. Another game, ``Aliens,'' employs pitch variations to inform players about the positions of targets, allowing for quicker navigation and interaction using arrow keys \cite{aliens_game}. Overall, sonification has proven to be a versatile tool in assisting BVI users, enhancing data analysis, accessibility, and user interaction through intuitive auditory cues.

\subsection{Image Generation and Editing with AI tools}
Generative AI has opened new avenues for creating accessible technology. Recent research has explored its use in making short-form videos accessible through video summaries, as demonstrated in projects like AVscript and ShortScribe ~\cite{Huh_2023,van2024making}. It has also been employed to enhance image generation processes (GenAssist) and to design effective tactile graphics using image-to-image generation techniques (Text to Haptics) ~\cite{huh2023genassist, Tanaka2023,Huh_2023}. Complementary tools such as BeMyEyes, Apple VoiceOver, Android Talkback, Chromebook Chromevox, and Braille displays further assist users in navigating and understanding various visual mediums ~\cite{eyes2020my,leporini2012interacting, singh2023towards}. An innovative approach is exemplified by WorldSmith, a multimodal tool that enables users to design and refine complex fictional worlds through layered editing and hierarchical compositions, incorporating a prompt-based model that integrates text, sketches, and region masks as inputs ~\cite{dang2023worldsmith}. Similarly, Crosspower explores leveraging language structure to facilitate graphic content creation, highlighting both the potential and limitations of language as a primary medium for image editing ~\cite{Xia2020CrosspowerBG}. While natural language processing offers precision, its inherent limitations underscore the need for tools that provide greater control and flexibility in visual content creation and editing for visually impaired users.

\section{Formative Study with Blind Visual Content Creators} 
To better understand existing visual content authoring workflows and associated challenges with representation, feedback, and iteration, we conducted semi-structured interviews with five blind experienced visual content creators.

\begin{table*}[ht]
    \centering
    \begin{tabular}{|p{2cm}|p{2cm}|p{2.5cm}|p{2.5cm}|p{3cm}|p{3cm}|}
        \hline
        \textbf{Participant ID} & \textbf{Eyesight} & \textbf{Image Editing Experience} & \textbf{Participated in Formative Study?} & \textbf{Participated in Design Study?} & \textbf{Participated in Final Evaluation Study?} \\
        \hline
        P1 & Blind & Experienced & yes & no & no \\ 
        P2 & Blind & Experienced  & yes & no & no \\ 
        P3 & Blind & Experienced  & yes & no & yes \\ 
        P4 & Low Vision & Experienced  & yes & no & no \\ 
        P5 & Blind & Experienced & yes & no & yes \\ 
        P6 & Blind & Novice & no & yes & yes \\ 
        P7 & Blind & Novice & no & yes & yes \\ 
        P8 & Low Vision & Novice & no & yes & no \\ 
        P9 & Low Vision & Novice & no & yes & no \\ 
        P10 & Low Vision & Novice & no & yes & yes \\ 
        P11 & Low Vision & Novice & no & yes & no \\ 
        P12 & Blind & Novice & no & no & yes \\ 
        P13 & Blind & Novice & no & no & yes \\ 
        P14 & Low Vision & Novice & no & no & yes \\ 
        \hline
    \end{tabular}
    \caption{Participant characteristics across the three studies (formative need finding, design feedback, and user evaluation study). \change{ ``Blind'' refers to participants who are totally blind, meaning they have no functional vision. ``Low Vision'' refers to participants who have some degree of vision impairment, typically using a combination of a screen magnifier and a screen reader. ``Past Experience'' refers to experience with image editing software using a screen reader or other accessibility tools. ``Experienced'' indicates extensive prior experience, while ``Novice'' indicates limited or no prior experience.}}
    \Description{This table summarizes the characteristics of participants across three different studies focused on formative need finding, design feedback, and user evaluation. It consists of six columns: Participant ID (P1 through P14), Eyesight status (Blind or Low Vision), Past Experience (Experienced or Novice), and three columns indicating participation in Interviews, Design Study, and Evaluation with 'yes' or 'no' responses. "Blind" refers to participants who are totally blind with no functional vision, while "Low Vision" refers to participants with partial vision impairment or related eyesight illnesses who typically use a combination of a screen magnifier and a screen reader. "Experienced" indicates prior experience with image editing using editing software or through SVG code using a screen reader or other accessibility tools. "Novice" indicates limited or no prior experience. Participants P1 to P5 are experienced and have participated in interviews, but their participation varies in design studies and evaluations. Participants P6 to P11 are a mix of blind or low vision novices, with varying participation in the design study and evaluation but no interviews. The last three participants (P12 to P14) are novices with different eyesight statuses and have only participated in the evaluation.}
    \label{tab:participant_details}
\end{table*}

\subsection{Participant Recruitment}
We recruited five participants by email using contacts in our immediate network. All participants self-identified as blind and had extensive experience authoring tactile graphics as part of their professional work or personal interest. Each semi-structured interview lasted between 45 and 60 minutes, and participants were compensated with a \$50 Amazon Gift Card for their time. The interviews were conducted over the Zoom video-conference platform~\cite{Zoom}. Table~\ref{tab:participant_details} provides a consolidated summary of all participants across different studies with specific columns indicating whether or not they were involved in each of the studies. Many of our participants engaged throughout the iterative design process. 

\subsection{Procedure}
Our semi-structured interviews consisted of three parts. In the first part, we asked participants questions about their visual content authoring workflow. Example questions include ``Can you tell us about the last time you authored a tactile graphic? What was the graphic about?'', ``Can you walk us through your process for authoring that tactile graphics using your current tools?'', ``Can you tell us about the tools you usually use and how they help you with the authoring process?'' Based on this context and common ground, we proceeded to ask them questions to understand current challenges and needs for visual content authoring tools. We asked, ``What are some pain points or challenges in your authoring workflow?'' and asked follow-up questions depending on their pain points. In the last phase of the interview, we asked participants about their experience of co-creating tactile graphics with AI, if any. We then brainstormed with the participants using guiding questions such as, ``What is the input you would initially provide to ChatGPT, and what output would you expect it to return?''
and ``how do you imagine iterating and editing on the graphics using ChatGPT?'' in order to understand their expectation and perceptions towards AI. All sessions were video recorded, and in a few instances, participants also emailed us artifacts of their creations.

\subsection{Analysis}
After the interviews, video recordings of the sessions were transcribed using the Zoom transcription features, and we read through the transcripts to ensure correctness (approximately 300 minutes of recordings). Using an open-coding approach~\cite{denzin2011sage}, two of the authors then independently coded all of the transcripts using ATLAS.ti~\cite{atlasti}. Over multiple cycles of discussion amongst the authors, we synthesized the key emergent themes around visual content authoring. We continued refining these themes until every category and subtopic was addressed, and no additional themes surfaced.  Specific themes included qualitative descriptions and characteristics of good tactile graphics, resources used to learn about visual content, tooling, color, visual attributes, specific features for editing, perception during authoring, limitations of current workflows, unexpected outputs from generative AI, and imagined uses.

\begin{figure*}[t!]
  \centering
  \includegraphics[width= \textwidth]{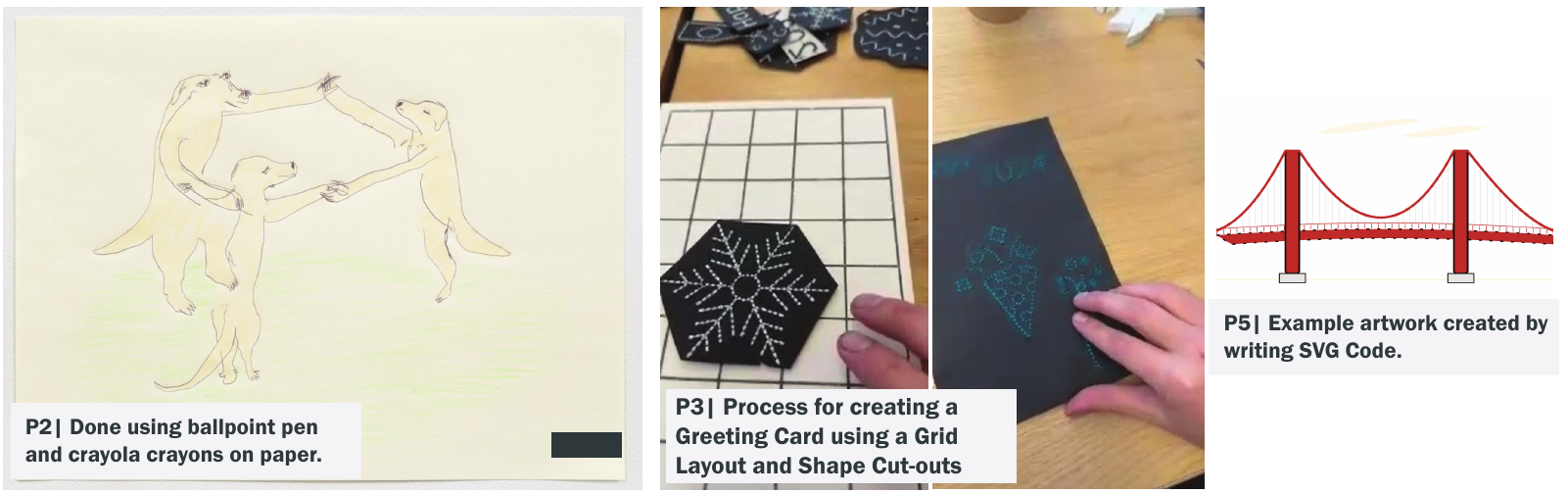}
  \caption{The participants' previous works, including a pen and crayon drawing, the original grid layout for placing images, and an artwork of a bridge written in SVG code. From the left, P2 showed us drawings using ballpoint pens and crayola crayons on paper. P3 showed us the process of using a grid layout to create shape cutouts. On the right, P5 showed us an example artwork of the Golden Gate Bridge using SVG Code.}
  \label{fig:formative}
  \Description{The participants' previous works including a pen and crayon drawing, the original grid layout for placing images, and an artwork of a bridge written in SVG code. From the left, P2 showed us drawings using ballpoint pens and crayola crayons on paper. P3 showed us the process of using a grid layout to create shape cutouts. On the right, P5 showed us an example artwork of the Golden Gate Bridge using SVG Code. From the right there is an image created by P2 | A sketched artwork of three figures, possibly a family, with two larger figures and a smaller one in the middle. They appear to be dancing or interacting playfully. The image is created with ballpoint pen and colored with Crayola crayons on paper. P3 | A photo of a hands-on crafting process showing a black greeting card with a cut-out pattern placed on a white grid layout. The card features a star-like pattern made from the cut-outs. P5 | A digital illustration of a red suspension bridge, possibly resembling the Golden Gate Bridge, against a white background. This artwork is noted as being created by writing SVG code.}
\end{figure*}

\subsection{Findings}
Across all sessions, participants reported needing to navigate multiple tools to author spatial content (both visual and tactile). Depending on the fidelity of the graphics they were creating, participants made use of low-fidelity tools, such as Wikki Stix and Sensational Blackboards. In many cases, they would later switch to more high-fidelity vector graphics authoring tools, but that often required assistance from sighted users.  When digitally authoring, participants reported frequently printing out the in-progress content as tactile graphics to assess and provide feedback or revise on their own.  Based on these authoring workflows, here we focus our findings on key low-level steps during authoring and associated challenges:

\subsubsection{Composition and Layout}
In many of the content that participants reported creating, they used existing graphic elements as a starting point, and the focus was more on composition. For instance, in one session, P3 designed holiday-themed greeting cards: \textit{``I was given a bunch of images, and I am using a physical grid to position them. I then work with a sighted individual to run it through Inkscape and print it out to create a holiday greeting card.''} Other participants also described using a similar grid-based approach for layout composition. In such cases, they reported needing to perform several calculations about the dimensions of the image and grid coordinates to place the items in the correct position. According to P5: \textit{``So okay, 50 units is half an inch. 25 units is a quarter of an inch, and with that kind of spatial reasoning and a lot of time in practice, you kind of get a sense of what the numbers you're dialing in are going to create on your canvas\ldots however, this becomes complicated to re-calibrate when editing the location many times.''} During composition, participants reported needing agency in specifying in detail where different elements will go. As P1 mentioned: \textit{``Say I am creating a seascape, in my mind, I can imagine how it will look, I know where everything goes, it is all in my head, this is what I want.''} Based on these insights, we infer that authoring tools that combine constructive and generative approaches should \textbf{provide precise control over the composition and layout of generated content without needing to manually calculate specific coordinates (D1)}.

Specific to using existing graphics, participants regularly looked for online content as a starting point to create their illustrations. According to P3: \textit{``my first process is to look for images online, and I'll look for keywords like black and white line art coloring pages and outline drawings.''} Further, P3 adds that they often need additional processing to make the visual compatible with tactile graphics: \textit{``I did re-scale and do some color management with a software called Tactile view to eliminate some unwanted color. And I use the latest version of Windows photos to zap some backgrounds that were creating distracting background dots''}. As an alternative, two participants also reported having learned how to code support vector graphics (SVG) manually to create digital content but the process can be cognitively intensive. P1 also added that they don't always want a hand-drawn look and feel. Therefore, we want tools \textbf{to make it easy for creators to access tactile graphics compatible illustrations (D2)}. 

\subsubsection{Iterative Editing}
Across sessions, participants mentioned a range of intents for editing the graphics being created. The majority of editing was done to resize elements. When digitally authoring, participants reported frequently printing out the in-progress graphics to assess the level of detail and whether elements were too close to each other, or if the element was too small to perceive the detail. In the seascape example, P1 mentioned: \textit{``We want it to be fun when we interact with it, feeling the curve of the shell or legs of the crab. So if the image is too small, those details get lost, and I have to make it bigger.''} Other edit operations included changing the appearance of the objects, such as curves and edges, or making some details less prominent through the use of primary and secondary lines to clarify which details should stand out. Based on these insights authoring tools should support \textbf{``dimension-based editing as well as feature-based editing (D3)''}

\subsubsection{Feedback During Authoring}
A key challenge in digital authoring tools is the lack of feedback during authoring. All participants reported seeking inputs from sighted individuals to assess and describe the graphic being authored. Alternatively, they have to print it out as tactile graphics to assess on their own. According to P5: \textit{``\ldots, she [sighted spouse] can look at it and tell me you know where to move it or if it needs to be adjusted. But if I'm by myself, then yeah, it's a lot of iteration, just a lot of printing back and forth.''} Participants aspired future tools would require fewer print cycles. Specific to editing tasks, participants commented that it would be helpful for tools to provide feedback on well-known constraints such as overlapping edges or extending beyond the border. According to P1: \textit{``have visual spell-checks\ldots have categories of things like what lines extending beyond borders.''} Participants also mentioned detailed verbal descriptions would be helpful (e.g., \textit{``on the top left corner you have a sea shell (P1)''}). Moreover, participants described such verbal descriptions should be collaborative and through a dialog where the creator can go back and forth to get into different kinds of details. Lastly, participants also reported that by nature, image editors rely on visual cues to understand the state of an image, making them inaccessible to those with visual impairments. It would be helpful if the tool had both verbal descriptions and sound feedback to convey the state of the graphics. Drawing from these insights, tools should \textbf{aim to minimize the number of print iterations (D4)} by \textbf{providing verbal and auditory feedback through dialogic interactions (D5).}

\subsubsection{Use of Generative AI}
Three of the five participants reported having explored different ways to use ChatGPT in their authoring workflow. For instance P3 reporting using GPT to get an initial SVG representation of an object and then iterated on their own. P1 was enthusiastic but also cautioned about its limitations based on their experience. According to P3: \textit{``It's fun and interesting to write a description and get a sense of what AI has created, but it is also challenging. It doesn't do a good job of spatially putting them where you want them to be. Thing A and Thing B and how you want them to be related to each other. Being able to specify that would be very helpful.''} They further added that describing takes a lot of work; it can be cognitively intensive and ambiguous. Along similar lines, P4 expressed that AI might help with creating some initial shapes to help guide their own authoring but not doing anything elaborate. Lastly, participants felt that if they were trying to create something unique, AI may not be able to accurately generate the illustration.

\section{User Experience}
\begin{figure*}[t!]
\includegraphics[width=\textwidth]{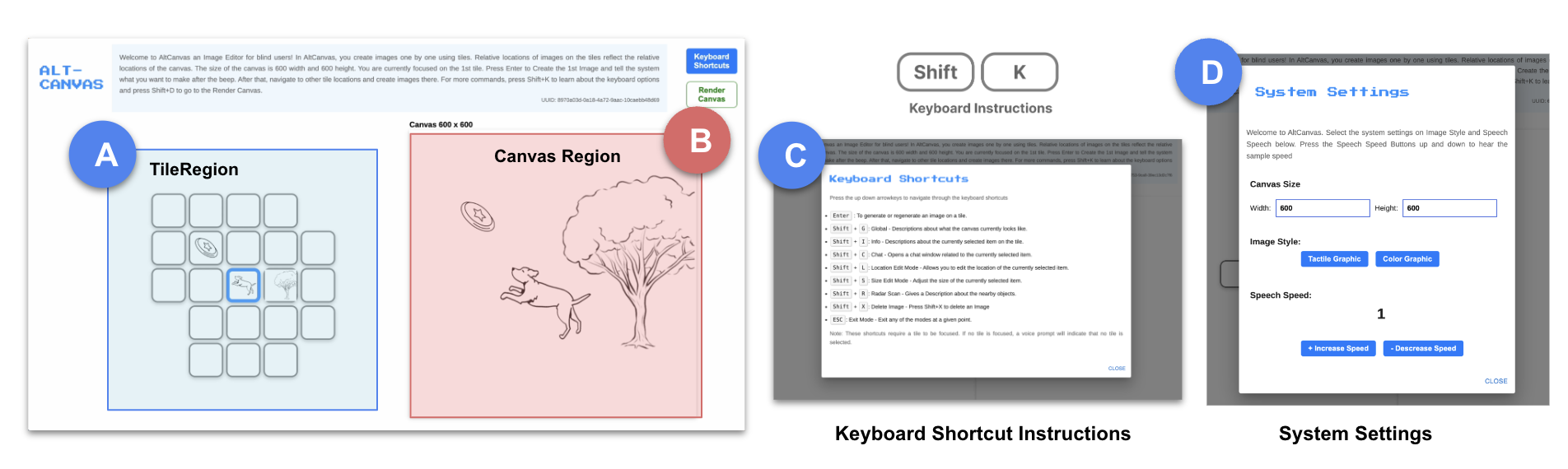}
    \caption{ \textbf{Main Interface} (A) Region of the Tile Based Interface. (B) Region of the Canvas Interface. (C) Region of the Keyboard shortcut commands (D) System Settings with Canvas Size, Image style, and Speech Speed. The screen (D) will pop up initially for the user to set settings. Users can access the keyboard commands screen C when pressing SHIFT + K. Users can navigate through the screen (A) and (B) regions while editing the image.}
    \label{fig:ux_onLoad}
    \Description{ A series of interface screenshots labeled from A to D illustrate different sections of the AltCanvas. A. Region of the Tile Based Interface. B. Region of the Canvas Interface. C. Region of the Keyboard shortcut commands D. System Settings with Image Style and Speech Speed. The screen D will pop up initially for the user for settings. Users can access the keyboard commands screen C when pressing SHIFT + K. Users can navigate through the screen A and B region while editing the image. In part A, titled "TileRegion" has a matrix of tiles, which are used to select and place images within the canvas. The tile currently selected is highlighted. Part B, called "Canvas Region," displays a canvas area with a 600x600 pixel grid where a line drawing of a running dog and a tree has been placed. Part C features a pop-up window with "Keyboard Instructions," listing shortcuts like 'Shift + K' for keyboard command instructions, and other combinations for various editing actions. Finally, part D shows the "System Settings" window, where users can adjust the canvas size width and height, image style between tactile and color graphics and control speech speed for audio feedback.}
\end{figure*}

Based on the guidelines from the formative study, we developed \system to help BVI users create visual content. As shown in Figure~\ref{fig:ux_onLoad}, \systems interface consists of two main regions: (1) on the left half of the interface is a tile view for authoring, and (2) on the right is an image view that will render the image being authored.

The tile view itself is comprised of \textit{dynamically expanding} interactive tiles that the user can navigate through directional commands (up, down, left, right). Each tile represents a single object on the image canvas. It provides a functional proxy for the creation and manipulation of that object, enabling users to control the object's properties effortlessly. The layout of the tiles on the tile view closely aligns with the relative positions of objects on the canvas. The tile view starts with a single tile, and once an object is associated with that tile, eight new tiles are added adjacent to that tile (left, right, top, down, top-left, top-right, bottom-left, bottom-right) to place objects relative to that tile object. Note that the tiles are all \textit{uniform} and do not represent the distance or size of the object on the canvas. Rather, they provide users with an easy way to navigate through images on a canvas and to select regions on the canvas for new generations by utilizing relative locations of image objects. In an earlier iteration, we experimented with absolute positioning based on the object arrangement on canvas by encoding size and distance (similar to the current grid-based techniques our participants reported). For instance, we can imagine dividing the canvas into a $5\times5$ grid and assigning one or more tiles to an object depending on how much space it occupies.  However, this made the navigation less consistent and required large changes in the tile positioning as objects moved around on the canvas. In the final design, we opted to encode just the relative position as tiles (based on design consideration D1). Size and distance are encoded using sonification and speech, which we will describe later in this section (based on design consideration D5). 

To better understand how \system supports a constructive and generative image authoring workflow, let us look at how Otto, a blind user, can create an artwork of his \textit{dog playing with a Frisbee in a park}. 

\subsection{Setup and Tutorial}
To begin with, Otto opens \system on his web browser. Because this is the first time Otto is using the system, he presses SHIFT + K to hear the keyboard commands in the system. This will open a popup with the keyboard commands listed. \change{These shortcuts were designed with reference to existing blind-accessible games and industry-defined accessibility practices, ensuring consistency and learnability (e.g., using SHIFT+ [LETTER] of the function name).} Using the arrow keys, Otto can navigate through the 10 commands in order (see Appendix for a list of keyboard shortcuts). The first down arrow key will read the first command SHIFT + G Global canvas description to Otto. Further, \system supports stereo audio with panning across the left and right microphone to support directional navigation on the canvas.

\subsubsection{System Setting: Speech Rate}
Since users have different speeds at which they recognize speech, depending on their proficiency with a screen reader, a system should allow the user to control different levels of speech rate. Before starting the prototype, users can choose their desired speech rate. Users can select from three distinct speech rate options within the system. The highest setting, rated as 3, matches the rapid delivery typical of screen readers. The intermediate option moderates the speed to a level between typical human speech and the swift pace of screen readers. The lowest setting, rated as 1, is calibrated to the comfortable listening speed for the everyday user.

\subsubsection{System Setting: Image Style}
As different users have varying needs for image creation with editors, our system offers primarily two methods of image authoring support: images to share with general audiences and tactile graphics. Users can select the type of image they wish to create—a colored version to share with general audiences or a tactile version that can assist with tactile graphic generation.

\subsubsection{System Setting: Canvas Size}
While the canvas size is set to default based on the user's screen width, on the system setting pop-up, users have the option to change the size of the canvas through keyboard input.

\subsection{Adding Objects to the Canvas}

Once familiar with the keyboard navigation commands, Otto exists the tutorial view and sees \systems the main authoring interface. By default, \system focuses on the single new tile on the tile view and greets Otto with an auditory guide: \textit{``You are currently focused on the first tile. Press Enter to generate the image on the 600 by 600 canvas.''} Otto presses Enter, and a distinct beep earcon confirms the system is ready for voice input. Following the beep, Otto voices his request, \textit{``Create an image of a dog.''} The system quickly processes his command and seeks confirmation, audibly prompting, \textit{``Detected: Create an image of a dog. Press Enter to confirm or the Escape key to cancel.''} Otto confirms by pressing Enter, signaling the system to generate the image. Based on design consideration D2, we use a text-to-image model to generate the image.

Once the image generation is complete, it is rendered on the canvas. By default, the first image is positioned at the center of the canvas. \system provided Otto with a detailed description of the generated image (D4, D5): \textit{``Dog has been generated. The coordinates of the image are 250 by 250. The dog is a golden retriever with golden hair and a smiling expression. The image measures 100 by 100.''} Initially, all images are generated at a size of 100 by 100. The coordinates of these images correspond to their relative positions on the canvas. Otto can retain this initial generation and proceed with further image creation and editing, or he can press the Enter key to regenerate the image with a different description or press Shift + X to delete the image if he wishes to try other objects for his artwork. 

Once the dog image is generated, 8 new tiles are added adjacent to the dog image tile. This allows Otto to begin adding other objects to the scene relative to the dog object. During keyboard navigation of the tiles, when Otto reaches the edge of the existing tiles, he will hear an audio cue, a ``thump'' sound indicating he has reached the edge of the tile blocks. On a current empty tile, Otto will hear a navigation sound depending on whether he is moving up, down, left, or right. Based on iterative feedback, users preferred different directional sounds as they were navigating the tile view. When Otto enters a tile with an image, he will hear the name of the image itself. With the initial image accepted, Otto can now continue authoring the scene. He decides to add more elements to his canvas. By focusing on the tile to the right of the dog tile, he voices another command,\textit{ ``Add an image of a tree with a thick trunk''}, and follows the same confirmation process as before. The system seamlessly integrates this new element into the existing canvas, maintaining spatial awareness and providing Otto with real-time updates.

\subsection{\change{Perception of AI Generated Content}}

\begin{figure*}[t!]
\includegraphics[width=\textwidth]{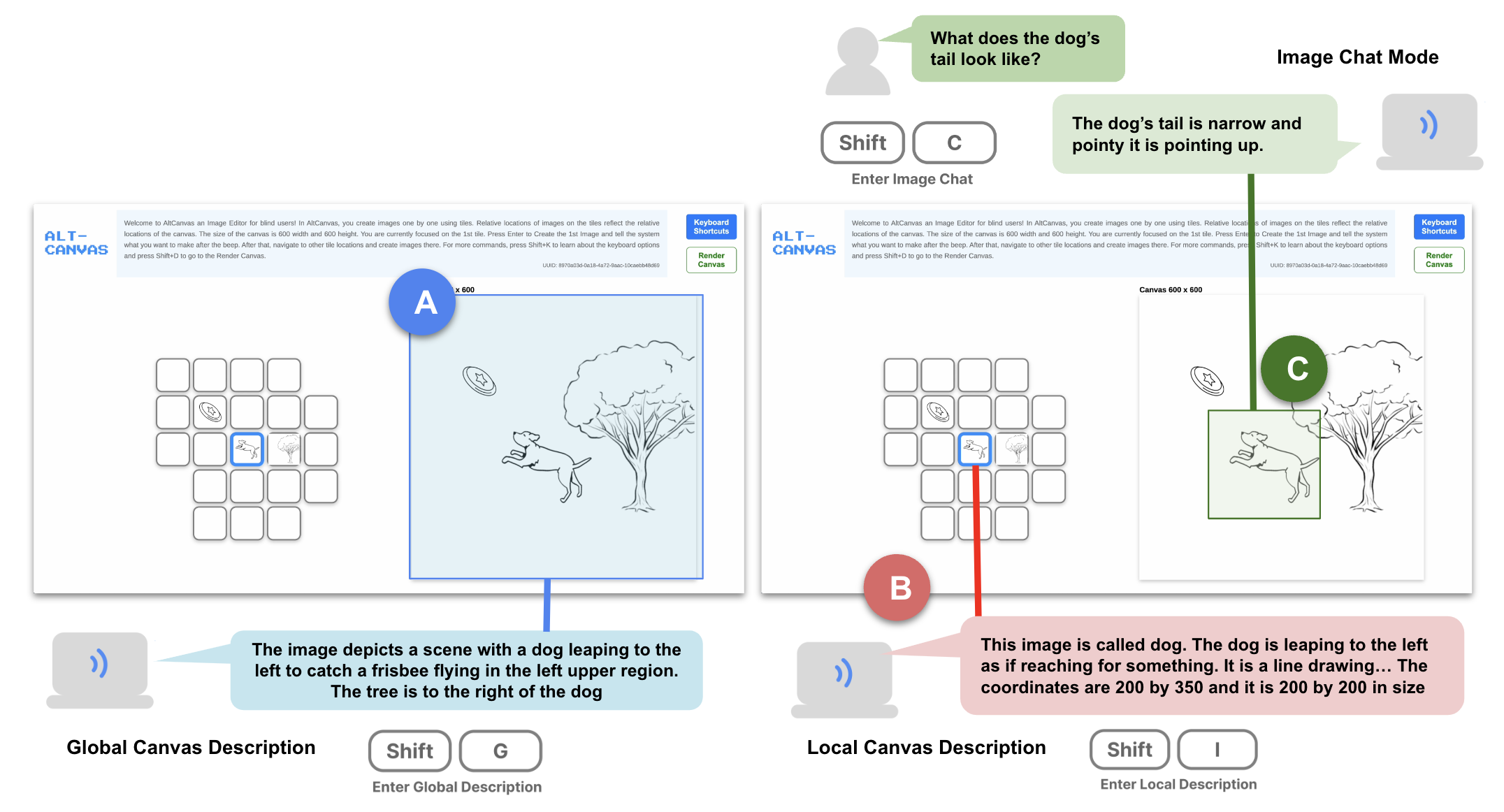}
    \caption{ \textbf{Image Descriptions} (A) Global Description of the canvas with multiple objects on it. The user can press the SHIFT +G command to hear the global description of the canvas. (B) Local Description: the user presses the SHIFT +I command to activate the local information. This will describe the image on the current tile to the user. (C) The Image Chat function. This command can be accessed through SHIFT +C. This feature will answer the question the user has about the image on the current tile to the user}
    \label{fig:ux_describe}
    \Description{ The two screenshots of interface to describe the AltCanvas's image descriptions. On the left, A. Global Description of the canvas is available. The user can press the SHIFT +G command to hear the global description of the canvas. On the right, B. Local Description activates the local information explanation by pressing the SHIFT +I command. This will describe the image on the current tile to the user. C. The Image Chat function. This command can be accessed through SHIFT +C. This feature will answer the question the user asks about the image on the current tile. In detail, the section A displays the 'Global Canvas Description' where a user has given a general description of a scene featuring a dog leaping towards the left to catch a frisbee, with a tree to the right of the dog. In section B, called 'Local Canvas Description,' there's a focused description of a particular element dubbed 'dog.' It describes the dog's action, the art style, and provides specific coordinates and size for the drawing on the canvas. Section C shows 'Image Chat Mode' where a voice command asking for the appearance of the dog’s tail is given, and the software responds that the tail is narrow, pointy, and pointing upwards.}
\end{figure*}

Now that Otto has generated a few images, he wishes to understand the layout and orientation of objects on the canvas. There are a total of four commands that Otto can use to understand the image: \textbf{Global Canvas Descriptions, Local Information, Radar Scan, and Chat }(to support design consideration D5). By using the SHIFT + G command, Otto activates the Global Canvas Description feature. This command provides an auditory overview of the entire canvas. The system describes the overall layout, any overlapping images, their relative positions, and the overall ambiance of the canvas, giving Otto a sense of visual aesthetics and composition. Once Otto presses this command, he hears the global description: \textit{``Your canvas currently features a golden retriever in the center with a medium-sized tree to its right. Both images are well-separated with no overlap, set against a clear background.''} At any time, Otto can press the Escape key to exit the narration. Second, the SHIFT + I command allows Otto to obtain detailed information about a specific image or tile he navigates to. When this command is activated, the system provides a description that includes the image's precise coordinates on the canvas and its current dimensions. This localized information helps Otto understand the specific details of individual elements within his artwork. Otto focuses on the dog tile and presses the SHIFT + I command. The system speaks: \textit{``The image depicts a golden retriever sitting upright and facing the viewer. The fur of the retriever is long, and it has a smiling face, with its tongue sticking out.''} 

Third, to get a sense of objects surrounding the active tile, Otto can activate the Radar Scan feature by pressing the SHIFT+R keys. Based on the user's current location, the radar scan identifies the nearest objects by name and measures the distance from the current image to others sequentially. For instance, pressing SHIFT+R while focusing on a dog would result in the scan announcing \textit{``Tree, 20 pixels away.''} Finally, for inquiries beyond the scope of automated descriptions, Otto can use the SHIFT + C command. This feature is designed for interactive engagement with the canvas (D5). Upon pressing SHIFT + C at an image tile, the system prompts, \textit{``Ask a question about the image and I will answer.''} Following a beep earcon, Otto can vocalize specific questions about the image. This could include inquiries about color, style, or any other detailed attributes not covered in standard descriptions. Otto asks: \textit{``Describe the shape of the dog's tail?''} and the system responds, \textit{``The dog's tail in this image is raised and curved upwards, reflecting a sense of excitement or alertness.''}

\subsection{Editing and Composition}

\begin{figure*}[t!]
\includegraphics[width=\textwidth]{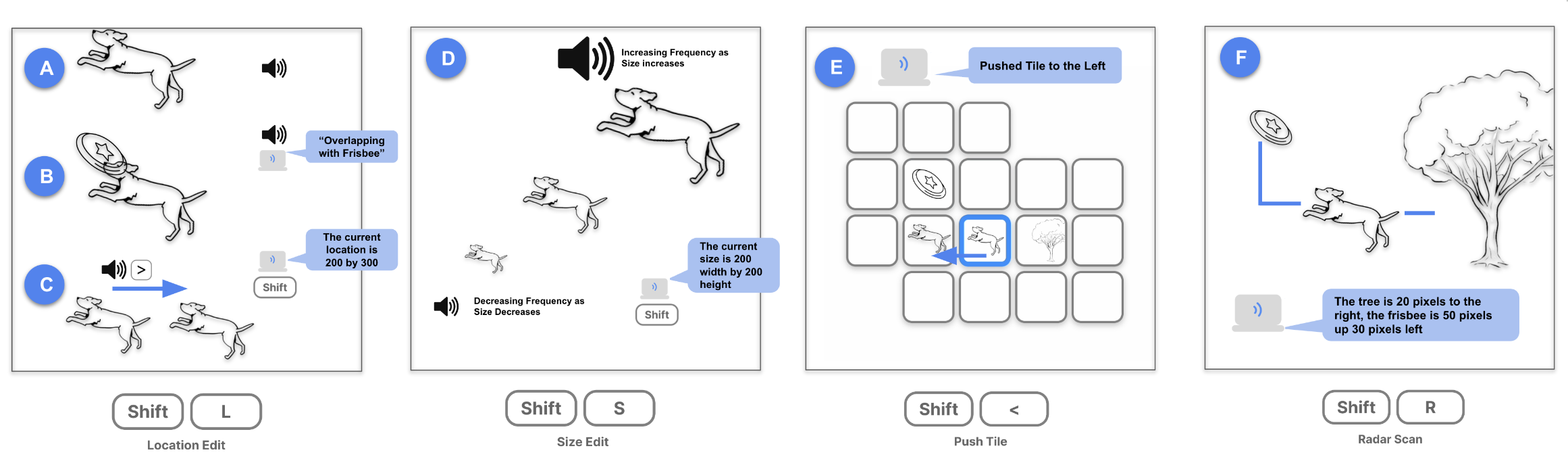}
    \caption{ \textbf{Image Editing Operations} Edit image locations with SHIFT + L and arrow keys: A thump sound indicates edge collision (A), speech notifies object collision (B), spatial sounds and coordinates describe movement (C). SHIFT + S adjusts size with variable frequency tones (D).  \change{When the size increases, the frequency increases, while as the size decreases, the frequency decreases.}   Tile manipulation and radar scan reveal layout and distances (E, F). }
    \label{fig:ux_interact}
    \Description{Edit image locations with SHIFT + L and arrow keys: A thump sound indicates edge collision (A), speech notifies object collision (B), spatial sounds and coordinates describe movement (C). SHIFT + S adjusts size with variable frequency tones (D). Tile manipulation and radar scan reveal layout and distances (E, F).
    Panel A features a simple line drawing of a dog in different positions, with a speech bubble indicating sound feedback when interacting with the image.
    Panel B depicts the same dog with a speech bubble saying "Overlapping with Frisbee" to signal a conflict in the image layout.
    Panel C shows an arrow pointing to the right with a speech bubble, and text indicating that the current location is 200 by 300, suggesting a feature to move elements on the canvas.
    Panel D provides audio feedback information where the frequency of the sound increases with the size of the image element, which in this case is a drawing of the dog. 
    \change{When the user presses the up arrow key to increase the size of an element, the system will play a sound with an increasing frequency to reflect the growth in size. Conversely, when the user presses the down arrow key to decrease the size of the element, the system will play a sound with a decreasing frequency to indicate the reduction in size.}
    Panel E has a grid similar to panel A, with a highlighted tile indicating that it has been pushed to the left, showing how to move tiles within the interface.
    Finally, panel F uses a radar scan metaphor to describe the spatial relationship between the elements on the canvas, with the tree being 20 pixels to the right, the frisbee 50 pixels up and 30 pixels left from the dog's position.}
\end{figure*}

Using the above perceptual feedback features, Otto can iteratively and flexibly edit images as he adds them to the canvas. \system provides Otto with four essential editing operations to refine his artistic creations. These include\textbf{ modifying the location and size of objects, pushing images around the canvas, and deleting unwanted elements.} These operations were most commonly described by the formative study participants to compose objects in the scene. We deliberately opted not to support fine-grain edits of individual objects through direct manipulation, such as enlarging the size of the dog's ear. Instead, we use generative features and natural language prompts for such modifications. In section~\ref{sec:discussion}, we discuss these design trade-offs. 

\subsubsection{Editing the Location of an Image}
To adjust the position of an image, Otto presses SHIFT + L to enter location edit mode. In this mode, using the arrow keys moves the image across the canvas. Each movement triggers an auditory notification confirming the action. If Otto's adjustments cause the image to overlap with another object or reach the canvas edge, he receives specific audio cues. These cues include warnings of overlaps (\textit{``Overlapping with Image X''}) and sounds indicating the edge of the canvas. To check the image's coordinates during editing, Otto can press SHIFT, which tells him the current position, with each press of an arrow key shifting the image by 20 units.

\subsubsection{Editing the Size of an Image} To modify an image's size, Otto initiates the process by pressing SHIFT + S, activating the size edit mode. Here, the UP and DOWN arrow keys adjust the image's size. Increases in size are accompanied by a rising earcon frequency, while decreases produce a lowering frequency. This sensory feedback helps Otto visualize the changes in real-time. If he needs to know the exact dimensions during resizing, pressing SHIFT provides this information. Adjustments are made in increments of 10 units per arrow key press while maintaining the overall aspect ratio of the generated image.

\subsubsection{Rearranging the Image}
When Otto needs to make space between images or reorganize the layout, he can use the push operation. By selecting an image and pressing SHIFT + ARROW KEY in the desired direction, the image shifts, clearing space for additional elements. This operation is confirmed audibly, \textit{``Pushed Image to the [direction],''} providing Otto clear feedback on his action. This allows Otto to add new images in between existing images (e.g., adding a Frisbee object between the dog and the tree). The tile view serves as a persistent reference to augment his spatial cognition as he makes the edits. 

\subsubsection{Deleting the Image} If an image no longer fits Otto's vision, he can remove it by pressing SHIFT + X. This command deletes the selected image, and Otto immediately hears, \textit{``Deleted Image on the tile.''} The removal of the image results in an empty tile, which alters the auditory navigation landscape, helping Otto understand that the space is now available for new creations.

\subsection{Rendering the Final Image}

\begin{figure*}[ht]
\includegraphics[width=\textwidth]{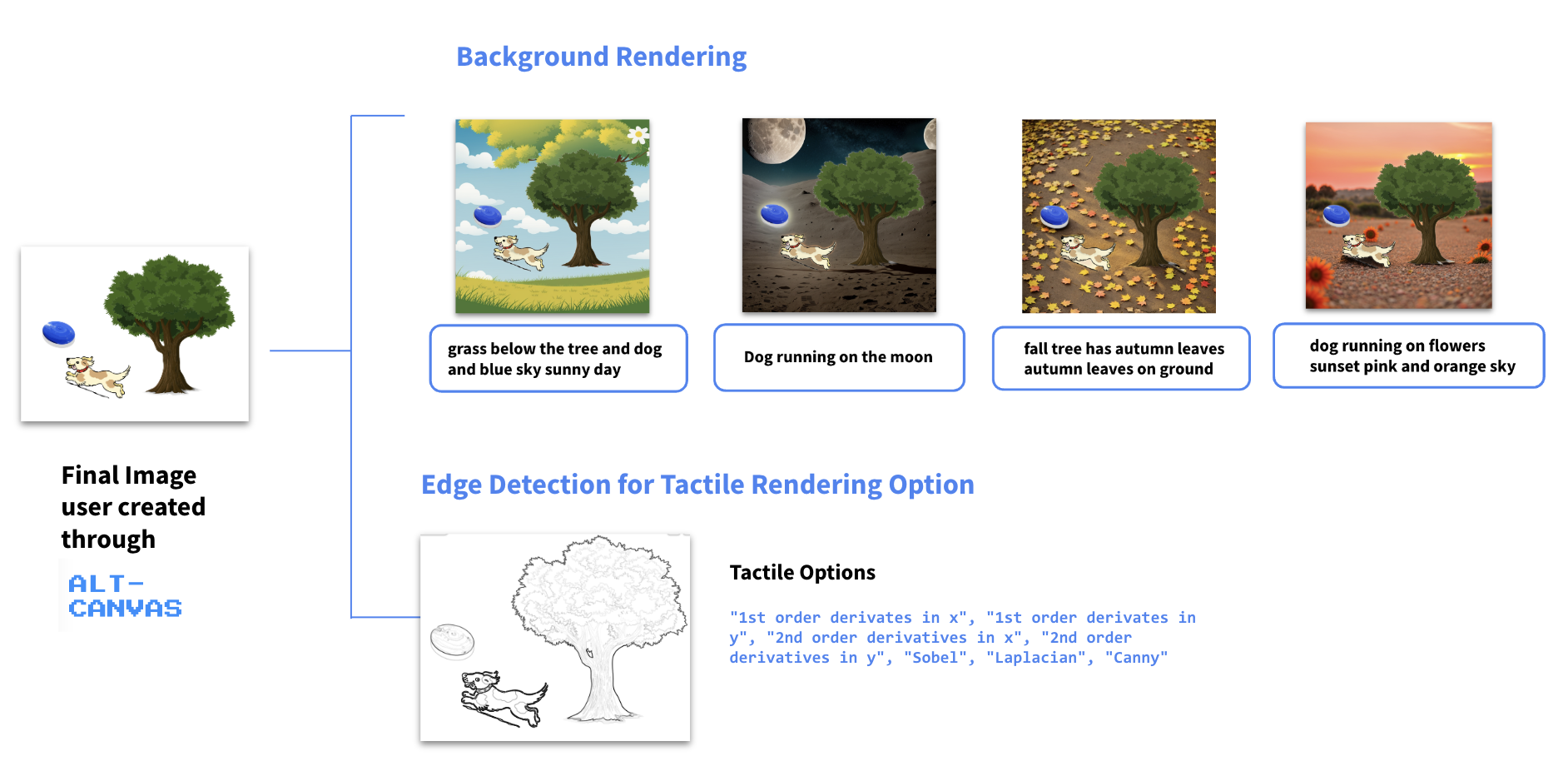}
    \caption{\textbf{Final Image Rendering} The figure illustrates the background rendering options and edge detection features in AltCanvas. Four different rendered backgrounds showcase how users can select various scenes for their final image composition. The second rendering operation illustrates the tactile rendering operations available to the user.}
    \label{fig:render}
    \Description{The figure shows a two-tiered illustration. Once the user has generated the final image through AltCanvas, the figure depicts the rendering options that are available to the user. The first option is, 'Background Rendering,' The images display four examples of a rendered scene featuring a dog, a tree, and a frisbee (the image the user has generated) in different environments: a sunny day with blue skies, a night scene with a moonlit backdrop, an autumn setting with leaves on the ground, and a dusk scene with a sunset. Each background has a the prompt that the user inputs to generate the final image. The lower tier, 'Edge Detection for Tactile Rendering Option,' presents a black and white outline of the same scene for tactile rendering purposes, with a list of technical options for edge detection algorithms including Sobel and Canny.}
\end{figure*}

After Otto completes editing his images on AltCanvas, he moves to the final stage, where the images are rendered into their finished forms. This section includes two specialized rendering options tailored to enhance tactile and visual experiences. For tactile graphic rendering, \system uses a dedicated model designed to optimize the image for tactile graphics. This rendering process adjusts the texture and relief of the image, making it suitable for tactile perception and interpretation. This option is particularly valuable for BVI users like Otto, as it transforms the digital image into a vector graphics format that can be printed using a tactile graphics printer. Additionally, Otto can choose to re-render the image to enhance its visual qualities, such as incorporating more naturalistic backgrounds or adjusting the color palette. This operation refines the visual elements of the image, making them more appealing and realistic. To do this, Otto provides a speech description of the type of background rendering he wishes to do on this final image. The re-rendering process might include adding shadow effects, lighting adjustments, and blending elements to create a cohesive scene that visually communicates Otto's artistic intent. This version can be shared with a sighted audience or embedded into other content, such as talk slides or blog articles on the web. Example results of tactile and color graphics using our \system are shown in Figure~\ref{fig:render}.

\section{System Implementation} \label{sec:system}
\begin{figure*}[ht]
\includegraphics[width=\textwidth]{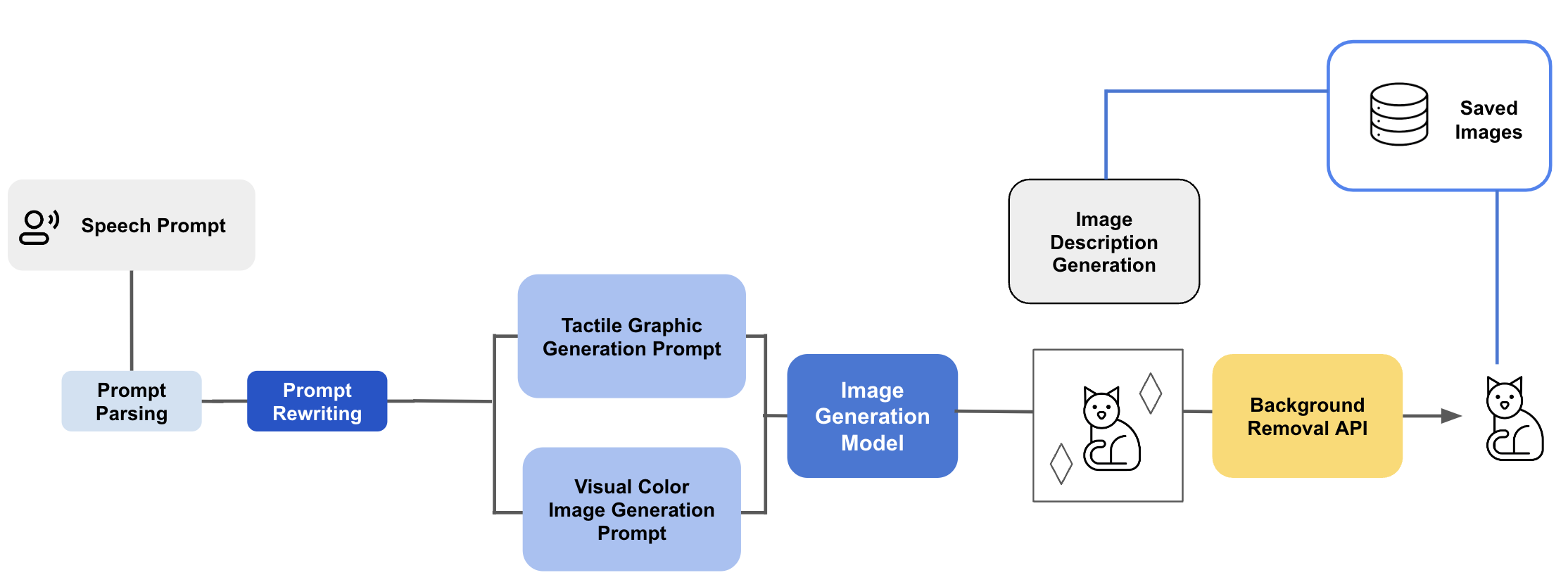}
    \caption{ \textbf{Image Generation Pipeline} The audio of the user is captured and parsed to single out the main components of the prompt using a wink-pos-tagger. The parsed prompt is processed into Tactile Graphic Generation and Visual Color Generation Prompts based on the type of image the user wishes to generate. This to final prompt is passed into the image generation model (model: gpt-4-turbo, parameters: n=1, style=natural, quality= hd) based on the initial prompt and image type requirements. The final image is passed through a background removal API and the url is used to generate descriptions of the given image. } 
    \label{fig:architecture}
    \Description{ Pipeline for Image Generation : The audio of the user is captured and parsed to single out the main components of the prompt. Theprompt is processed into Tactile Graphic Generation and Visual Color Generation Prompts based on the type of image the user wishes to generate. This final prompt is passed into the image generation model and goes through a verification process based on the initial prompt and image type requirements. The final image is passed through a background removal API and the url is used to generate descriptions of the given image. The image is a flowchart detailing the process of generating an image from an audio recording. The flow starts with an audio recording, which goes through 'Prompt \change{Rewriting', where we make the user prompt command more comprehensive to create detailed graphics.} 'Tactile Graphic Generation Prompt' and 'Visual Color Image Generation Prompt,' both leading into an 'Image Generation Model.' The output of the model is an 'Image Description Generation,' which presumably describes the generated image. This description is linked to 'Saved Images,' suggesting that the generated images are stored. Additionally, there is a 'Background Removal API' that processes the saved images to remove their backgrounds, illustrated by a transition from a full image to one where only the subject remains.}
\end{figure*}

\system is a web-based application employing a client-server architecture. Figure~\ref{fig:architecture} illustrates the input-output pipeline, highlighting the system's two primary components: the image generation module and the image description module. The process begins when a user issues a speech command, which is then parsed to create an image prompt for the Large Language Model (LLM) image generation model. Following image creation, the image description module generates a comprehensive, accessible description of the generated image.

\subsection{Image Generation}
\change{The user's speech input (e.g., \textit{``create an image of a cat''}) is initially transcribed and then passed along to a prompt rewriting pipeline (see Appendix section~\ref{sec:prompts}). For tactile graphics generation, our base prompts align with the Braille Authority of North America (BANA) guidelines ~\cite{BANATactileGraphics2022}, emphasizing key features such as the absence of perspective, clear outlines, simplification, and elimination of unnecessary details. For other image types, our prompt focuses on limiting generation to a single object and excluding text. Once generated, the image undergoes background removal. The resulting image is then placed on the canvas. This refined prompt is passed to the GPT-4o model (parameters: n=1, style=natural, quality=hd) for image generation. Due to model limitations in infographic generation, our system currently focuses on object images.}

\subsection{Image Descriptions}
After image generation, the resulting URL is input into the GPT-4o model to create a description tailored for visually impaired users. When the global canvas description mode is activated, the entire canvas is captured and processed through the GPT-4 API to generate a comprehensive description. The canvas's HTML element is captured using the html2canvas library and rendered as an image before being sent to the system for description generation~\cite{html2canvas}. Users can employ this feature during the editing process to understand the current canvas state, checking for image overlap and positioning.  The prompt used for description generation is provided in the Appendix (section~\ref{sec:prompts}). During location and size editing, users require more rapid system responses for efficient manipulation.

\change{\subsection{Image Editing}}

\subsubsection{Size Editing}
\change{Sound frequency maps to object size - higher frequencies for larger objects, lower for smaller ones as preferred by participants. Each time the user presses the up arrow key, the size of the object increases by a specified amount maintaining height and width ratio. Users can press the SHIFT+I key after pressing the up arrow key to understand the change in the size of the object. It will give them verbal feedback ``size 30'' (height 30, width 30).}

\subsubsection{Location Editing}
\change{The location of the object can be adjusted using the keyboard arrows. Users can move elements in four directions: up, down, left, and right. Spatial sonification provides audio cues as the object is moved - users will hear an up, down, left, and right spatial sound as they move around the objects. If a user bumps into another element, they hear a thump sound indicating the collision. Users can press the SHIFT+I key to check the object's current location in coordinates (X,Y) after moving it. The centerpoint of the object is called.}

\subsubsection{Radar Scan}
\change{Radar Scan calculates the distance between the current image that the user has selected and the surrounding elements. The center point of the object is used to determine the direction the object is located in. The descriptions include both proximity and direction information (e.g., ``The tree is 20 pixels to the right, the frisbee is 50 pixels up and 30 pixels left'').}

\subsubsection{Dynamic Tiles}
\change{Tiles and images maintain direct correspondence, reflecting changes in positioning during editing. If an element is moved on the canvas, the corresponding tile's position will be updated to match this new layout.}

\subsection{Implementation Details}
The web-based client is developed using React and hosted on AWS Amplify \cite{awsamplify}. It incorporates the Web Speech API for speech recognition and utilizes the SpeechSynthesisUtterance interface for audio feedback to users\cite{mdnSpeechSynthesisUtterance}. An initial onboarding section guides users through various audio checks, facilitating interaction with spatial audio generated by Tone.js for navigation in four directions (left, right, top, and bottom) \cite{tonejs}. Audio files, primarily MP3s, are loaded from an AWS S3 bucket and played using Tone.js's spatial audio capabilities. \systems backend server is implemented using Node.js and Express.js \cite{AmazonS3, expressjs}. It integrates API calls to the OpenAI API and a Background Removal API (removal.ai) for enhanced functionality  \cite{openaiapi, removalai}. Additionally, the wink-pos-tagger library was used to parse the initial user prompt before sending it to the model \cite{winkpostagger}. Sound effects for sonification, stored as MP3 files in a public AWS S3 bucket, were sourced from Pixabay under an open-access agreement. The backend infrastructure is hosted on AWS Elastic Beanstalk with a Caddy reverse proxy \cite{pixabay, caddyserver}. Images generated are transiently handled, only being stored on the server if the user opts to save them in the final Render Canvas page. To render the final image, we finetuned a model that does tactile image conversion. This model takes in an image that is a non-tactile graphic and converts it into a tactile graphic by stripping out color and creating more pronounced outlines. To render the final image, we used the stable-diffusion model.

\section{Iterative Design Feedback}

\begin{figure*}[t!]
\includegraphics[width=0.8\textwidth]{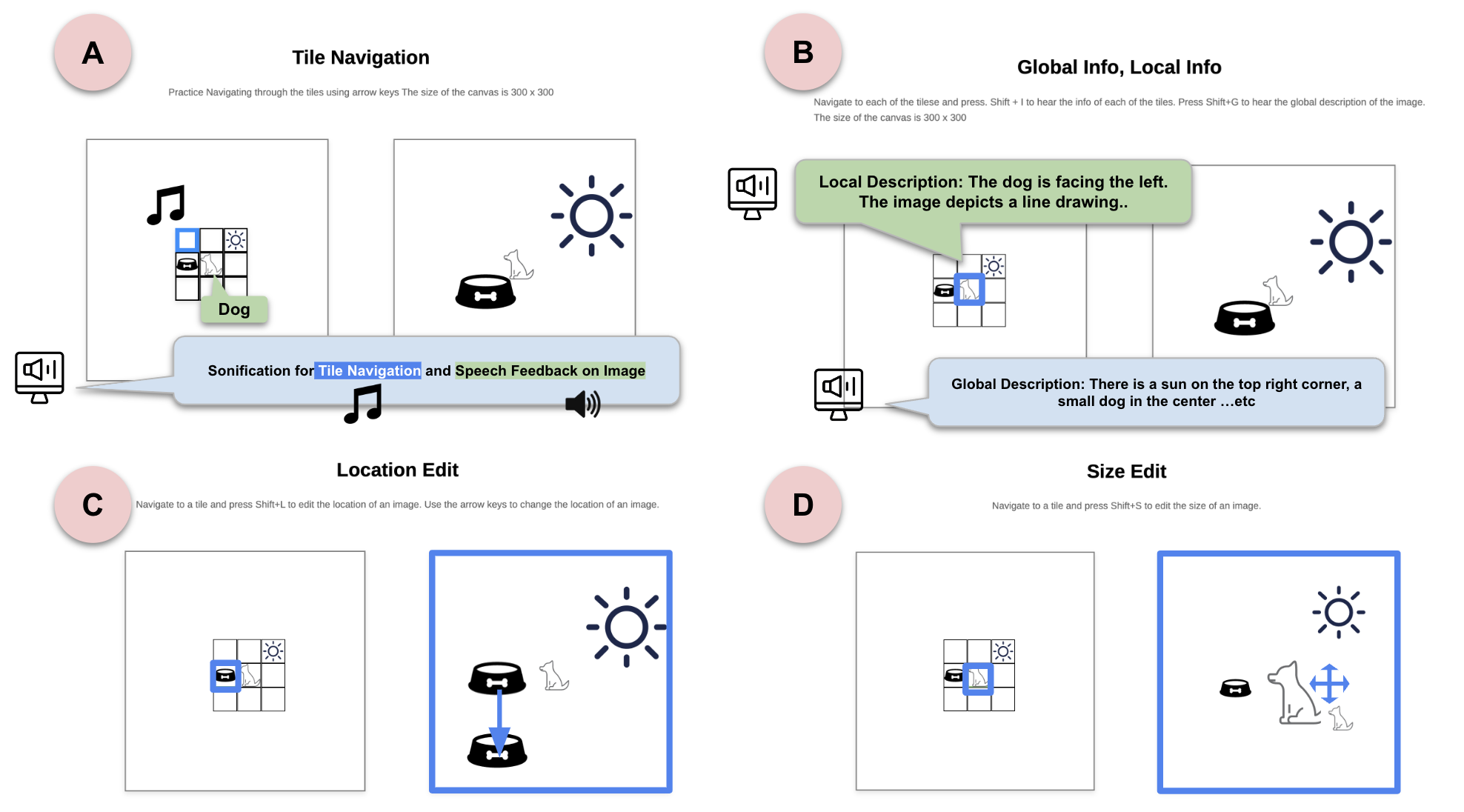}
    \caption{ \textbf{Overview of Interactions Reviewed in Preliminary Study for Design Feedback} The four main types of interactions that were tested in the pilot study. (A) Tile Navigation: here, we tested if users were able to gain a spatial understanding of the canvas through navigating through the tiles. (B) Global Info, Local Info, here we tested the descriptions of the image to gain an understanding of the different types of descriptions wanted. (C) Location Edit: Here, we studied how users navigated through changing the location of images and the types of information they wanted. (D) Size Edit: here, we studied how users changed the size of objects on the canvas.}
    \label{fig:pilot}
    \Description{ The main four types of interactions that were tested in the pilot study. (A)Tile Navigation: here, we tested if users were able to gain a spatial understanding of the canvas by navigating through the tiles. (B) Global Info, Local Info, here we tested the descriptions of the image to gain an understanding of the different types of descriptions wanted. (C) Location Edit: Here, we studied how users navigated through changing the location of images and the types of information they wanted. (D) Size Edit, here, we studied how users changed the size of objects on the canvas. Panel A is titled ``Tile Navigation'' and illustrates a grid where each tile represents an element of the image. The panel shows a music note and a sun icon, with a label ``Dog'' on one of the tiles. A speech bubble indicates that sonification is used for tile navigation and that speech feedback is provided on the image content.
    Panel B is labeled ``Global Info, Local Info'' and depicts two speech bubbles: one for a local description stating ``The dog is facing the left. The image depicts a line drawing.'' and another for a global description mentioning a sun in the top right corner and a small dog in the center.
    Panel C, ``Location Edit,'' shows an interface where the position of image elements can be edited. A selected tile with a dog is highlighted, and a keyboard shortcut is suggested for editing the location.
    Panel D, ``Size Edit,'' demonstrates how the size of an image element can be adjusted. It shows a tile with a sun, and an arrow icon implies the scaling function, with a keyboard shortcut for size editing mentioned.}
\end{figure*}

To design \systems features, we conducted a preliminary design study with six participants (2 blind and 4 with low vision). Using early working prototypes of \system design alternatives, we aimed to gather feedback on the intuitiveness of a tile-based interaction interface and to explore the intricacies of speech recognition, sound models, and sonification techniques. Concretely, we evaluated four categories of interactions. : (1) Tile Navigation: to understand whether or not the tile interaction was intuitive to users, (2) Global Info, Local Info: to understand if image descriptions were sufficient (3) Location Edit: To understand the location edit details of moving objects (4) Size Edit: To understand the size edit details of changing around sizes of objects (see Figure~\ref{fig:pilot}). Participants were recruited by reaching out to local organizations that work with BVI individuals, including the Vista Center for the Blind~\footnote{https://vistacenter.org/}, and LightHouse for the Blind and Visually Impaired~\footnote{https://lighthouse-sf.org/}.  All sessions were conducted over Zoom using the web-based prototypes. The sessions lasted between 30-45 minutes and participants received \$25 for the time. Details about the participants can be found in Table~\ref{tab:participant_details}. Here, we report the specific tasks along with summaries of participant feedback, which we incorporated into the final system design and implementation. 

\subsection{Interaction Alternatives and Participant Feedback}
For the study, we started with a simple illustration that we created that comprised a dog, a dog bowl, and the sun. A total of three objects in the scene. 

\subsubsection{Tile Navigation} To assess our tile-based interactions with authoring, we instructed participants to explore the scene through the tiles with an arrow key and, at the end, provide us a description of the location of objects on the canvas. As the users moved through the tiles, they were able to hear sonified feedback and image descriptions when they landed on objects. For instance, the dog tile would make a subtle barking sound, the bowl the sound of metal, and the sun a `flare' sound. Further, we added a short navigation sound as feedback, indicating the focus was on the new tile. 

\textbf{Feedback from Participants:} All participants achieved a comprehensive understanding of the spatial orientation of the canvas via the tile-based interface. Each participant could accurately articulate the orientation of objects on the canvas, although the identical sounds used for different directional interactions (top, down, bottom, up) caused some confusion despite spatial audio enhancements. P4 suggested, \textit{``Sounds are like colors to us; I want to hear more dynamic and varied sounds for each direction.''} The majority favored hearing the direct names of objects on each tile rather than personalized sonified notes. P2 noted that unique sounds for each object might lead to cognitive overload due to the mental effort needed to link distinct sounds with specific objects.

\subsubsection{Global and Local Information} Secondly, we wanted to understand users' ability to process and utilize the descriptive information presented globally (whole system overview) versus locally (detailed segment information). Our goal was to determine the most effective method for conveying complex information to BVI users, which could significantly enhance their authoring and editing experience. In our prototype, users pressed the SHIFT + G and SHIFT + I commands to activate the global and local information and gave feedback on the descriptions they heard. The global description gave a description of the layout of the dog, the dog bowl, and the sun. The system said  \textit{``There is a dog in the center of the canvas, a dog bowl to the left of the dog, and then a sun on top of the dog.''} For the local description, the system says, \textit{``This is a line drawing of a dog. The dog is facing the left.'' }

\textbf{Feedback from Participants:}  In the descriptions, users found it essential to have both coordinate information and visual location descriptions available. They requested more detailed global descriptions to better understand the overlapping and relative positions of objects. Additionally, there was a call for more interactive features to query information about the canvas that was not covered in the descriptions, such as asking, \textit{``What color is the dog?''} or \textit{``What does the dog bowl look like?''} For the local information, users requested an accessible way to easily retrieve coordinate or size information quickly without having to go through the entire image description.

\subsubsection{Location Edit}
We gathered feedback on interactions related to editing and repositioning graphical elements within the interface. We aimed to assess the intuitiveness, accuracy, and responsiveness of the system when users engage in such modifications, which are critical for ensuring effective interaction. Users pressed the SHIFT + L command to activate the location edit mode, once they were focused on the object they wished to edit. Then, they used the arrow keys to change the locations of objects. Users heard speech feedback to know that they entered the mode and heard either verbal descriptions of object movement such as ``up 10'' or sonification earcons. For this task, we asked participants to move the bowl closer to the bottom edge of the screen. 

\textbf{Feedback from Participants:}  Through the use of spatial audio, users were introduced to a variety of earcons that effectively communicated changes in location. The initial earcon consisted of a simple chime that emitted a spatial sound with each keystroke. This was followed by direct speech feedback, which provided precise information on their movements across the canvas—for instance, ``up 10.'' Users commented on the lag of the speech feedback of the up 10 and how it could cause more confusion when the keyboard was pressed multiple times. Users commented that if they had a way to quickly know the changes of movement, the simple chime earcon would be sufficient. Overall, during location edits, users primarily sought two pieces of information: (1) the exact canvas location of objects and (2) the magnitude of their movements. For users adept with the grid coordinate system, navigating this setup was intuitive and straightforward. In contrast, those less familiar with the coordinates preferred a broader description of the object’s position, such as \textit{``slightly to the center-right.''} Moreover, users effectively discerned bump sounds at the edges of the canvas and identified sounds indicative of overlapping objects.

\subsubsection{Size Edit}
The final task is on how users adjust the size of visual elements within the interface. We wanted to understand the level of control and ease with which users can customize the visual aspects of the interface to suit their individual needs and preferences. Users pressed the SHIFT +S command to activate the size edit mode, once they were focused on the object they wished to edit. Then, they used the up and down arrow keys to change the sizes of objects on the canvas. Users heard speech feedback to know that they entered the mode, and heard a mixture of ``increase 10'' size feedback or increase size frequency sound feedback. For this task, we asked users to increase the size of the dog to make it larger than the bowl. 

\textbf{Feedback from Participants:} Contrary to typical size increase feedback, users identified an increase in frequency as an indication of larger object sizes, while a decrease in frequency signaled a reduction in size. As with location edits, users sought precise numerical details regarding size changes. Additionally, they expressed interest in understanding the relative sizes of objects on the canvas, achieved by activating the global information mode (SHIFT + G).

\section{Usability Evaluation}
We conducted a user study to evaluate the overall usability and effectiveness of our tile-based paradigm and the combined constructive and generative authoring workflow. The study was conducted via Zoom with 8 BLV users (6 blind, 2 low vision) as indicated in Table~\ref{tab:participant_details}. In addition to participants in prior studies, we recruited participants from local community organizations, including the Vista Center for the Blind~\footnote{https://vistacenter.org/}, and LightHouse for the Blind and Visually Impaired~\footnote{https://lighthouse-sf.org/}. The study took approximately 90 minutes for each one-on-one session, and participants were compensated with a \$50 dollar gift card for their time. 

\subsection{Procedure}

\subsubsection{System Checks (5 minutes)} At the start of each session, the participant conducted a system check using a calibration page in our system to confirm that the audio and speech were working in their system. This page was compatible with a screen reader, and participants went through the process on their own. For participants who were not familiar with their screen readers, we helped them navigate through this system and check through the remote control of the screen through Zoom. Participants tested their sound by pressing the left, right, and top buttons to hear the sounds that will be used in \system. The audio was tested through a practice round of screen recording. This process helped participants turn on browser settings for instances where the audio was not activated.

\subsubsection{Onboarding Tutorial (20 minutes)} 
Once the audio setup was complete, we gave participants a detailed hands-on tutorial to help them gain familiarity with \systems features. We first explained the Tile Navigation and Image generation process and asked them to generate an image of a spoon. We then proceeded to go through the image understanding and editing commands. Participants were instructed to create an image of a fork. Next, we helped them understand how to edit the location of images using the Location Edit (SHIFT +L) command. Using this, the participants were instructed to move the fork to the left of the spoon, thereby experiencing the overlap interaction feedback. Then, we asked them to use Size Edit mode (SHIFT + S) to experience increasing and decreasing the size of the image. Finally, participants tested deleting the Image (SHIFT +X) and pushing the image on a tile to create tile space (SHIFT + ARROWKEY)

\subsubsection{Study Tasks (45 minutes)}
Once participants indicated familiarity, they proceeded to work on three illustration tasks. We provided specific scene descriptions for the first two tasks with varying complexity, and the third task was open-ended. In \textbf{Task 1}, we asked participants to (1) Create an image of a dog. Make the dog above 150 pixels in size. (2) Create an image of a dog bowl. Move the dog bowl to the left of the dog and make sure they do not overlap. (3) Create an image of a clock above the dog and place it on the top of the canvas. \textbf{Task 2} required them to generate five objects and perform eight different edit interactions. Specifically, participants were instructed to: (1) Create an image of a dining table. Make the table above 200 pixels in size. Place it on the bottom of the Canvas. (2) Create an image of a potted plant. Place it on top of the table.  (3) Create an image of a window. Place the window in the top left corner. Make the window above 150 pixels in size. (4) Create an image of a clock. Place the clock in the top right corner.  Make the clock above 150 in size. (5) Create an image of a cat. Place the cat on the bottom left corner of the canvas. In \textbf{Task 3}, an open-ended task, we asked participants to create an illustration of their choosing without any guidelines. We collected system log data, including all of the prompts and actions taken for each task for each participant. Refer to Appendix X to see some samples of the generated images across all tasks.

\subsubsection{Feedback Elicitation (20 minutes)}
After the study, participants were asked to respond to a set of nine usability questions on a scale of 1-7. Then we followed up with an open-ended discussion with some initial guiding prompts:  ``What do you think of using this tool for authoring illustrations? '', ``How does \system compare to other tools or your conventional ways of making illustrations? '', ``Are there features that are missing or what would like to see in future iterations? Do you have any feedback to improve its usefulness?'' ``How do you envision that this tool can be useful in your daily life?'' 

\subsubsection{Evaluation of Created Illustrations}
After the final study, we printed out the graphics created by each participant and mailed it to them for feedback. We printed the graphics on a Swell Paper and used a PIAF printer to generate the tactile graphics. We added descriptions to each of the tactile graphics about the images that the graphic included. We followed up with them over email with questions to evaluate their printed graphics. Questions included ``Does the printed tactile graphic align with what you think you created in our tool?'',  ``If not, what are the specific misalignments between the generated image and your perception of using our tool?'', ``what information is currently missing about the image generation but could have been helpful during authoring and editing?'', ``what do you think you want to change about the images?'', and ``do you have any feedback to improve its usefulness?'' Participants responded to these questions by email.

\subsection{Results}

\begin{figure*}[t!]
\includegraphics[width=\textwidth]{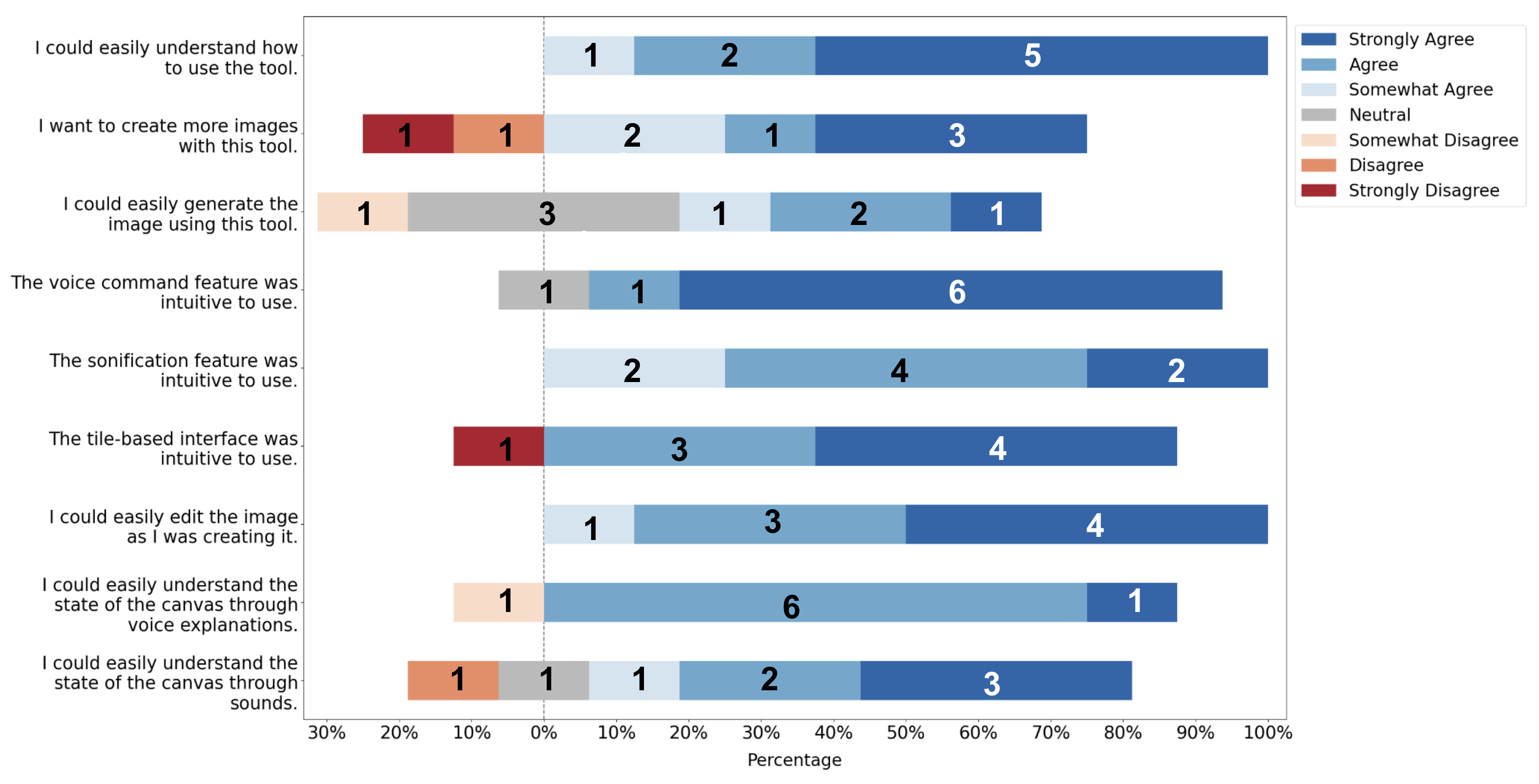}
    \caption{ \textbf{LIKERT Scale responses for the final study} LIKERT Scale responses for the final study on a 7-point scale. Highest satisfaction can be observed for the questions ``I could easily understand how to use this tool'' (avg = 6.5, std= 0.75), ``The voice command feature was intuitive to use'' (avg = 6.5, std=1.06). ``The tile-based interface was intuitive to use'' (avg= 5.8, std=2.03) and ``I could easily edit the image as I was creating it''(avg= 6.3, std=0.7). Mixed responses were received for ``I want to create more images with this tool''(avg=5, std= 2.3) and ``I could easily generate the image using this tool'' (avg= 4.8, std=1.3).}
    \label{fig:final_likert}
    \Description{ The image is a bar chart displaying user feedback on a tool, with each statement rated on a Likert scale ranging from ``Strongly Disagree'' to ``Strongly Agree.'' The vertical y-axis lists user statements, and the horizontal x-axis represents the percentage of responses. Darker shades of blue indicate stronger agreement, while red tones indicate disagreement. The statements, from top to bottom, are: I could easily understand how to use the tool. I want to create more images with this tool. I could easily generate the image using this tool. The voice command feature was intuitive to use. The sonification feature was intuitive to use. The tile-based interface was intuitive to use. I could easily edit the image as I was creating it. I could easily understand the state of the canvas through voice explanations. I could easily understand the state of the canvas through sounds. LIKERT Scale responses for the final study on a 7 point scale. Highest satisfaction can be observed for the questions ``I could easily understand how to use this tool'' (avg = 6.5, std= 0.75), ``The voice command feature was intuitve to use'' (avg = 6.5, std=1.06). ``The tile based interface was intuitve to use'' (avg= 5.8, std=2.03) and ``I could easily edit the image as I was creating it''(avg= 6.3, std=0.7). Mixed responses were received for ``I want to create more images with this tool''(avg=5, std= 2.3) and ``I could easily generate the image using this tool'' (avg= 4.8, std=1.3). The primary reason behind this is because of the unreliability of image generations from the AI model. The majority strongly agree that they could easily understand how to use the tool and that the voice command feature was intuitive. Most users agree or strongly agree that they want to create more images with the tool. Opinions vary more on the tile-based interface and the ease of editing images, with a spread across neutral to agree.}
\end{figure*}

\subsubsection{Image Generations}
Regarding the ease of use of the voice command feature, users responded to the question ``The voice command feature was intuitive to use'' with an overall high satisfaction ($avg=6.5$) (Figure~\ref{fig:final_likert}). Responses to the statement, ``I want to create more images with this tool'' were varied ($avg=5$, $std=2.3$) (Figure~\ref{fig:final_likert}). P10 responded that they wanted to further explore ``How creative this could become.'' The feedback regarding image generation with the model was generally positive, though there were some limitations. To the question, ``I could easily generate images with this tool,'' 6 users agreed, while 2 users disagreed ($avg=4.8$, $std=1.3$) (Figure~\ref{fig:final_likert}). Most users were pleased with the quality of the generated images. The model often produced creative results from simple prompts, such as ``an image of a dog,'' leading to variations like ``beagle,'' ``golden retriever,'' or ``cocker spaniel.'' This unpredictability added a touch of fun for some users, with P7 and P5 commenting, ``Really fun, you never know what you'll get out of the model.'' and ``Easy to create another image, and it's fun to see what the AI creates.'' However, a few users felt frustrated when they couldn't generate exactly what they wanted. For P5, who had experience in generating SVG graphics, the lack of control over model responses was a source of concern. The participant commented that they had ``limited control over the creativity of the tool'' compared to generating free-form graphics on their own through SVG code. Despite this, the participant commented ``I was impressed with the owl and the coffee mug, and for blind folks who may not know how to draw certain items, this could be a good way to help educate them and build up their image context and knowledge.''

\subsubsection{Image Descriptions}
\begin{figure*}[ht]
\includegraphics[width=\textwidth]{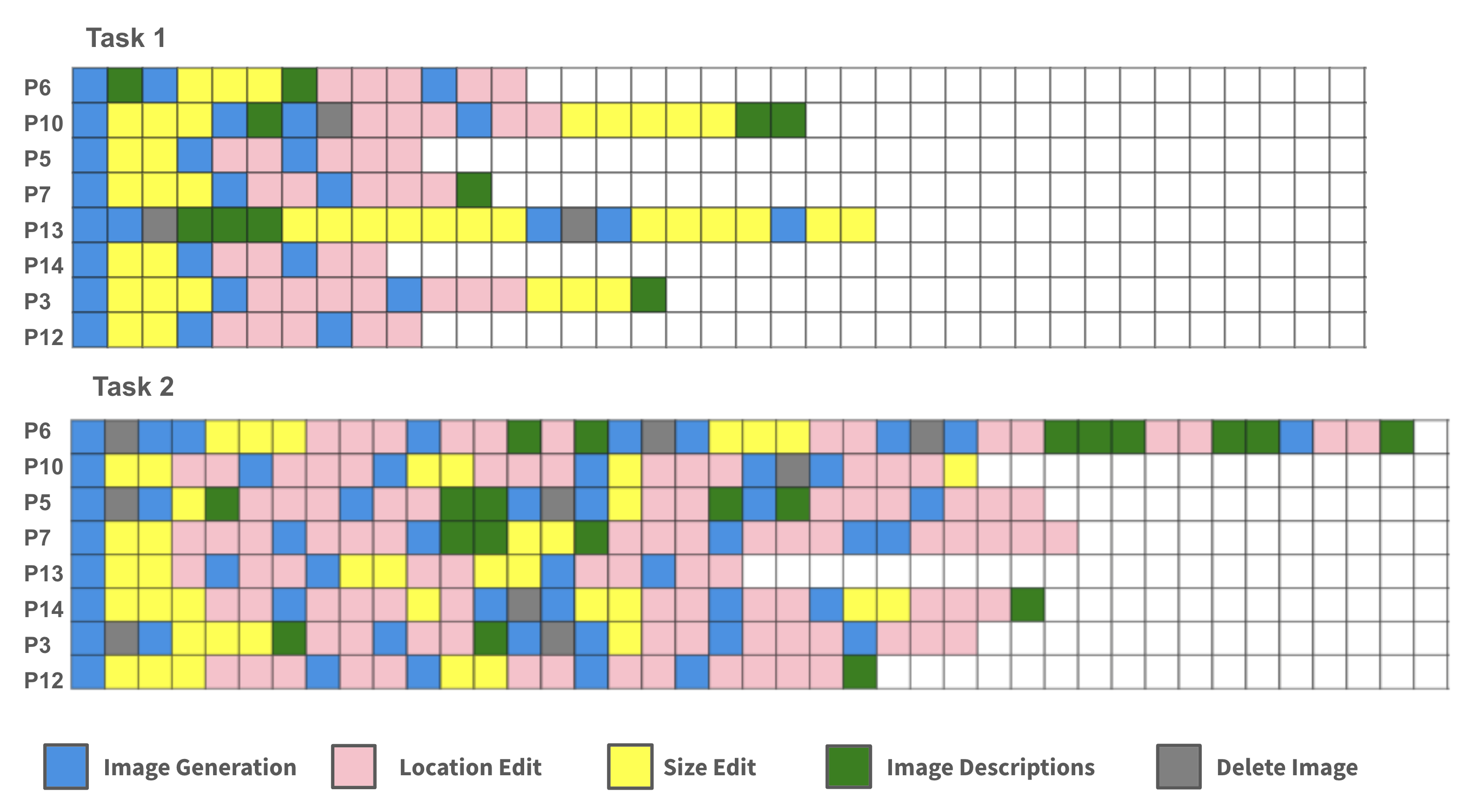}
    \caption{ \textbf{Sequence of Edit Interactions} Sequence of Edit Interactions for Task One, Simple Image and Task Two, Difficult Image. The boxes correspond to the editing durations. A general trend of global and local image descriptions used after multiple edit interactions and towards the end of image editing can be observed. The green image descriptions refer to the separate image description commands accessed by SHIFT +G (Global), SHIFT+I (Information), and SHIFT+C (Chat). }
    \label{fig:final_interactions}
    \Description{ The image is a grid-based chart representing the actions taken by participants in two separate tasks. Each row corresponds to a participant labeled P3, P5, P6, P7, P10, P12, P13, and P14. The columns represent different actions: Image Generation (blue), Location Edit (pink), Size Edit (yellow), Image Descriptions (green), and Delete Image (gray). Task 1 and Task 2 are shown in two separate sections of the grid. The participants' actions are marked on the grid with colored squares corresponding to the actions' legend, illustrating the sequence and frequency of each action performed by the participants across the tasks.A visualized Sequence of Edit Interactions for three tasks. The blue refers to image generation tasks, the pink refers to location edits, yellow refers to size edits, green refers to image descriptions, and gray refers to delete image. The boxes correspond to the editing durations. A general trend of global and local image descriptions used after multiple edit interactions and towards the end of image editing can be observed. For task one, a total of three images were generated, whereas in task two, a total of five images have been generated. Image generations are followed by an image delete for incorrect generations. The green image descriptions refer to the separate image description commands accessed by SHIFT +G (Global), SHIFT+I (Information), and SHIFT+C (Chat). The general trend shown in the chart is that participants engage in various image editing activities, with some actions being more common than others. In both Task 1 and Task 2, 'Image Generation' (blue) and 'Location Edit' (pink) are quite prevalent across most participants, indicating these are frequent steps in the tasks. 'Size Edit' (yellow) also appears regularly, but less often than image generation or location editing. 'Image Descriptions' (green) are used moderately throughout, and 'Delete Image' (gray) is the least frequent action, suggesting that participants do not often delete images once they have been generated or edited. The pattern of actions seems consistent between Task 1 and Task 2, indicating similar behavior across tasks by the participants. }
\end{figure*}

Users were overall satisfied with the generated image descriptions. To the question, ``I could easily understand the state of the canvas through voice explanations.'' users responded positively ($avg=6$, $std=0.5$) (Figure~\ref{fig:final_likert}). Through the interaction data, we observed users utilizing the image description commands accessed by SHIFT +G (Global), SHIFT+I (Information), and SHIFT+C (Chat) after consecutive edit interactions and towards the end of image editing (Figure~\ref{fig:final_interactions}). Image descriptions were also accessed by users during system edits as found in the Pilot Study. As users were editing the location and size, they often referred back to the coordinates and current width and height information to confirm their generations using the SHIFT key. Participants commented on certain confusions with the location coordinates and ``having to do my own math.'' We note that there is a trade-off with qualitative descriptors such as ``top-left'' and exact coordinates when participants desire precision. In future iterations, we aim to look at using the questioning features to help with computation. Additionally, users utilized the image descriptions after image generations to confirm the visuals that they had created.

\subsubsection{Image Editing Interactions}

\textbf{Tile-based system as an effective tool for relative image location navigation:}
Overall, users expressed satisfaction with the tile-based editing interface. To the question ``The tile-based interface was intuitive to use.'' participants responded positively ($avg= 5.8$, $std=2.03$) (Fig.~\ref{fig:final_likert}). P14 commented that the tile interface ``helps with figuring out spatial awareness.'' By navigating to tile locations with images, participants liked being able to locate the tiles on the interface and then move onto the canvas for further edits.
On the system's intuitiveness, ``I could easily understand how to use the tool.'' Participants responded that ($avg=6.5$, $std=0.75$). Despite the number of keyboard commands, users were able to easily navigate to image generation, location edits, and size edit commands throughout the completion of the tasks. 

\change{It took an average of 10.4 minutes (ranging from 5 to 22.83 minutes) to complete Task 1 and an average of 21.5 minutes (ranging from 13 to 28 minutes) to complete Task 2. For the freeform task, we gave a time limit of 15 minutes. We also calculated the average time users spent on generation and editing operations across sessions. On average, users spent 6 minutes and 28 seconds ($std = 3:01$) on image generation. Specific to editing, for size editing, the average total time was 5 minutes and 39 seconds ($std = 3:01$). The location edit action had an average total time of 6 minutes and 19 seconds ($std = 1:38$).}

\textbf{Sonification for Edit Interactions:}
To the question, ``I could easily understand the state of the canvas through sounds'' (to gather feedback on sonification), users responded ($avg=6$, $std=0.75$). P10 mentioned that the sonification feature in tile navigation and location edits ``helps with figuring out spatial awareness'' during the image editing. P3 commented that ``up and down sounds were intuitive for the tile navigation.'' Participants were satisfied with the sonification changes as image sizes increased in frequency when the image size increased and decreased when the size was decreasing. Users also expressed satisfaction with the Canvas Edge and Image overlapping sounds and were able to complete the overlap interactions (placing a potted plant on a table) and placing the window and clock successfully on the top corners of the canvas as seen in the Appendix.

\textbf{Verbal Feedback for Editing:}
Overall, participants effectively used verbal descriptions and image information throughout the editing process. The image descriptions were utilized to confirm image generations, verify current canvas states, and ensure that their desired interactions were displayed correctly (Fig.~\ref{fig:final_interactions}). For example, participants frequently used the SHIFT +G Global description to hear the global description of the canvas after placing the potted plant on the table. While participants had heard that the image was overlapping to confirm the final image, they listened to the global description before confirming completion.

\subsubsection{Quality of Final Illustration}
Five out of eight participants responded by email, offering feedback on the printed tactile graphics that were shipped to them. All participants responded positively to the match between their perceived illustration (using feedback from the tool) during the authoring and the final printed output. P3 responded that ``almost precisely how I imagined it would be placed.'' Participants also suggested additional features that could enhance the tool's functionality, such as ``options to know the location of the picture through degrees like around the clock—2 o'clock, 3 o'clock,'' and ``options to add backgrounds and adjust them with the editing tool.''

\change{Participants recommended several enhancements to improve both functionality and accessibility. They suggested expanding editing capabilities by adding background adjustment, image rotation/flipping, and multi-edit functionality. To increase accessibility, they proposed improving assistive technology integration, particularly enhancing compatibility with screen readers. Optimizing keyboard accessibility through streamlined navigation and editing shortcuts was also emphasized. Additionally, participants recommended implementing stereo audio spaces to facilitate more intuitive spatial navigation of the canvas. These proposed improvements aim to create a more comprehensive and accessible tool catering to users with varying levels of expertise and visual abilities.} 
\section{Discussion}\label{sec:discussion}

\system introduces a novel workflow for creating visual content through generative AI, offering enhanced control over editing interactions and scene composition. Our study reveals that users value control over their creative outputs and understanding of the underlying processes. \systems tile-based paradigm, serving as an alternative view of the drawing canvas, provides visually impaired users with diverse creative editing options, including dimension and location edits, image regeneration, and compositions. This approach offers a level of controllability previously unavailable, enabling a more engaged and informed creative experience.

\subsection{Broader Utility of \system}
\change{More generally, \systems tile-based paradigm can support spatial understanding and manipulation of a variety of visual content. Participants in our study, including those with no prior image editing experience, identified various potential applications for \system. These ranged from editing children's books to creating visual graphics and developing educational materials. Notably, participants suggested integrating the tile-based editing approach into mainstream software like Microsoft PowerPoint and Apple's publishing tools, indicating its potential to enhance spatial content editing for a wider range of users. }

\change{In enabling access to such applications, \systems tile-based interaction offers an alternative to traditional linear text descriptions and aligns more closely with natural spatial cognition. For instance, in educational settings, the spatial exploration facilitated by our tile navigation could enhance students' understanding of complex diagrams. Rather than relying on sequential descriptions like ``the heart is above the stomach, which is left of the liver,'' students using \system can navigate the tile grid using directional commands, building a more intuitive mental map. They might start at the heart tile, move down to the stomach, and then right to the liver, experiencing spatial relationships directly. This non-linear exploration allows for immediate comparison of element positions and sizes, fostering a more comprehensive understanding of the overall layout. }

\change{Additionally, \systems sonification features, such as distinct sounds for tile navigation and object overlap, provide multi-sensory feedback that further reinforces spatial awareness, offering a richer, more interactive learning experience than traditional text-to-speech descriptions. \change{For visually impaired users, sonification features serve a role similar to colors in visual imagery. Sonification provides real-time auditory feedback, creating aural landmarks that aid in forming mental maps of canvas layouts. Keyboard controls with specific sound feedback reinforce spatial changes, allowing users to discern object positions and relationships through varied tones. For instance, in music production, the tile-based system could be adapted for spatial audio mixing, with sound sources placed and manipulated on a virtual soundstage. For urban planning, sonification could represent traffic flow or population density, providing an intuitive way to understand city dynamics. Our study found that sonification effectively communicated spatial information and canvas state, providing an enjoyable editing experience. Users could adjust image locations using spatial sounds and detect overlaps through audio cues. This approach addresses the challenge of alternating between task interfaces, screen readers, and additional plugins often required for spatial navigation~\cite{Tutoria11y}.} }

\change{The broader implications of \system extend to the field of accessible technology and AI-assisted creative tools.  For instance, in professional contexts such as graphic design, \systems approach can enable visually impaired designers to create and edit complex layouts independently. In educational publishing, it could allow visually impaired educators to generate and arrange visual elements in presentations or digital textbooks. For game development, this approach could enable the creation of audio-centric games with rich, spatially-aware environments.  These examples illustrate how \systems integration of generative AI with intuitive interaction methods points towards more inclusive and adaptable creative tools, potentially transforming workflows and expanding creative possibilities across various professional contexts and disciplines.}

\subsection{Limitations and Future Work}

\subsubsection{Complex Editing Operations}
In designing \system, our primary focus was to utilize blind spatial cognition, allowing for a streamlined and iterative authoring process without adding complexity. Therefore, our current implementation does not support advanced editing features like one may find in line-by-line drawing tools or vector graphics authoring tools such as Adobe Illustrator. In our earlier iterations, we prototyped nested tile views at the object level (tiles representing parts of a dog) and scene level inspired by the nested grid drawing method by Kamel and Landay~\cite{kamel2000study}. However, since the tile views are configured dynamically, we found that the effort required to acquire the spatial representations of generated elements can become challenging for users to remember. Similarly, for our tile-based authoring, we opted for eight directions for relative object placement as opposed to fine-grained navigation, such as segments of 10-degree angles. Our directional implementation, while allowing for ``put it there'' spatial cognition, the degree of freedom can be constrained, requiring multiple steps to position objects. Future work can look at offloading more editing functionality to generative AI while exploring better and more accurate perceptual feedback. 

\subsubsection{Sonification Experiences}
Our implementation is limited in the range of expressive sonification interactions. Future work should look into generating more diverse and interactive sounds for editing interactions. Based on participant feedback, there was a desire to represent the entire canvas in stereo space. Future developments should focus on creating more diverse and interactive sounds for editing interactions. This direction in development would not only enhance the functionality of the tools but also significantly improve the user experience by providing a more intuitive and immersive editing environment.

\subsubsection{Use of Generative AI}
\change{While generative AI offers powerful capabilities for image creation, it also presents significant limitations in the context of illustration tools for visually impaired users. The unpredictability of AI-generated outputs can be challenging, as users may not always obtain the precise image they envision. This lack of fine-grained control over specific details can be particularly frustrating for users who have a clear mental image of their desired illustration. These constraints highlight the need for continued research into more controllable and interpretable AI models for image generation, as well as the development of hybrid approaches that combine AI generation with more traditional, user-controlled editing techniques. For instance, by incorporating features like user image uploading or Retrieval Augmented Generation (RAG), we can provide more control over image generation. Image uploading would allow users to work with existing visuals, potentially enhancing or modifying them within the \system environment. This could be particularly useful for tasks like adapting educational materials or refining pre-existing designs. RAG, on the other hand, could improve the accuracy and relevance of generated images by leveraging a curated knowledge base. For instance, when creating illustrations for specific subjects or styles, RAG could ensure that generated content aligns more closely with user intent by drawing from relevant visual and contextual information.}

\section{Conclusion}

In conclusion, our work presents a novel visual content authoring paradigm for BVI users, enhancing their ability to create, interact with, and understand visual content. By introducing a tile-based interface that provides an alternative view of the drawing canvas, we enable spatial comprehension and interactions for authoring visual scenes. Additionally, the use of generative text-to-image AI models allows for the generation of complex images from simple voice commands and natural language descriptions, providing a way for creative expression previously dominated by basic, less expressive authoring tools. Our approach combines the expressiveness of generative AI with a constructive method that respects user agency, enabling users to incrementally build and modify visual scenes with heightened control and flexibility. Throughout this project, our engagements with 14 BVI participants—through interviews, feedback sessions, and usability evaluations—have been crucial.  The findings from our evaluation show that participants were successfully able to create illustrations using \systems tile-based paradigm and authoring workflow. Our work contributes to a more inclusive environment for visual arts and digital content creation with generative AI.

\begin{acks}
We thank our participants for their time and valuable inputs in the design of \system. We also thank the reviewers for their feedback on the paper. This work is supported through the AI Research Institutes program by the National Science Foundation and the Institute of Education Sciences, U.S. Department of Education through Award $\#2229873$ - National AI Institute for Exceptional Education. 
\end{acks}

\balance{}

\bibliographystyle{ACM-Reference-Format}
\bibliography{99_refs}

\clearpage
\appendix
\appendix
\onecolumn
\section{Appendix}

\subsection{Keyboard Shortcuts}

\begin{table*}[ht]
\centering
\begin{tabular}{|c|c|c|c|}
\hline
\textbf{Keyboard Commands} & \textbf{Function} & \textbf{Main Interaction} \\ \hline
ENTER & Image Generation / Regeneration & Generate a new image with speech \\ \hline
SHIFT + G & Global Canvas Description & Hear global description about the canvas  \\ \hline
SHIFT + I & Local Image Description & Hear local description about the image on tile  \\ \hline
SHIFT + R & Radar Scan for Surrounding Objects & Hear the name of the object and numerical distance  \\ \hline
SHIFT + C & Image Chat & Ask a question about the image \\ \hline
SHIFT + L & Location Edit & Edit the location of an image  \\ \hline
SHIFT + S & Size Edit & Increase / Decrease the size of an image  \\ \hline
SHIFT + Arrow Key & Push Image & Push image tiles to create tile space   \\ \hline
SHIFT + X  & Delete Image & Delete an Image on the tile   \\ \hline
Arrow Keys & Tile Navigation, Location/Size Edit & Navigate through different spaces on tiles  \\ \hline
ESC & Quit / Stop & Exit an editing model, stop model speech \\ \hline
\end{tabular}
\caption{The 10 main keyboard commands that are used in this system. This table lists the 10 main keyboard commands used in a system and their respective functions and interactions. It has three columns: `Keyboard Commands', `Function', and `Main Interaction'. For example, pressing `ENTER' generates a new image with speech, while 'SHIFT + G' provides a global description of the canvas. Local descriptions, radar scans for nearby objects, image chats, and editing image location or size are performed with various `SHIFT' plus a letter key combinations. The `SHIFT + Arrow Key' pushes image tiles, 'SHIFT + X' deletes an image, and the arrow keys alone navigate through tiles. 'ESC' is used to quit or stop actions within the system. Each command aligns with a specific interaction for image editing and navigation within the system, aimed at enhancing user experience.}
\Description{This table lists the 10 main keyboard commands used in a system and their respective functions and interactions. It has three columns: 'Keyboard Commands', 'Function', and 'Main Interaction'. For example, pressing 'ENTER' generates a new image with speech, while 'SHIFT + G' provides a global description of the canvas. Local descriptions, radar scans for nearby objects, image chats, and editing image location or size are performed with various 'SHIFT' plus a letter key combinations. The 'SHIFT + Arrow Key' pushes image tiles, 'SHIFT + X' deletes an image, and the arrow keys alone navigate through tiles. 'ESC' is used to quit or stop actions within the system. Each command aligns with a specific interaction for image editing and navigation within the system, aimed at enhancing user experience.}
\label{tab:keyboard_command}
\end{table*}

\subsection{TaskOne: Sample User Graphic Generations}
\label{sec:user_graphic}
\includegraphics[width=0.33\textwidth]{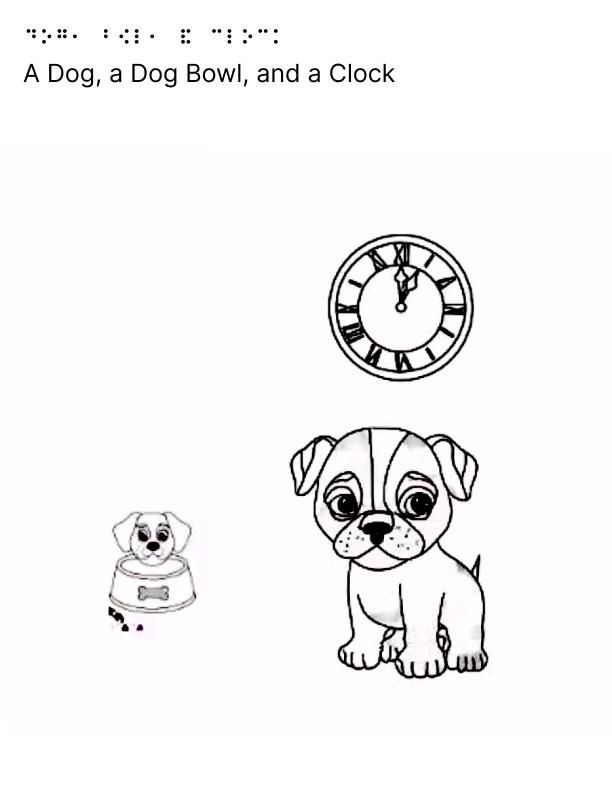}
\includegraphics[width=0.33\textwidth]{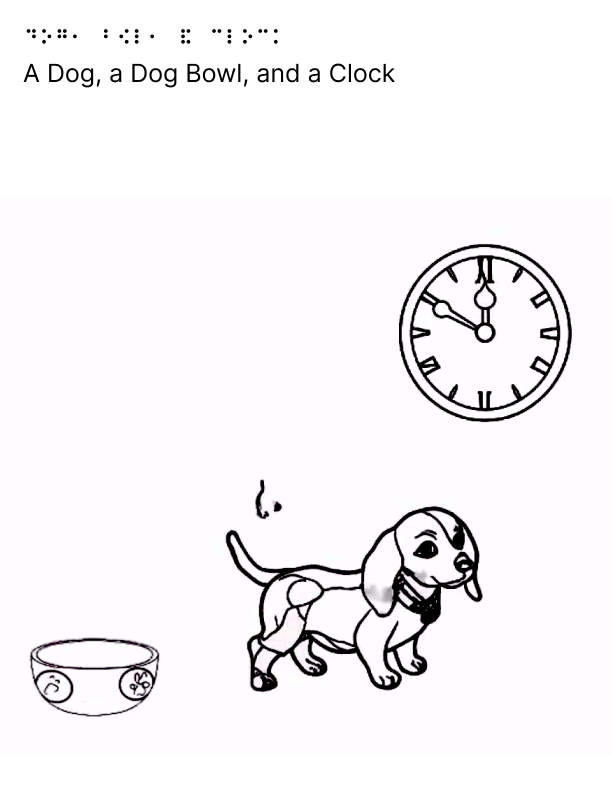}
\includegraphics[width=0.33\textwidth]{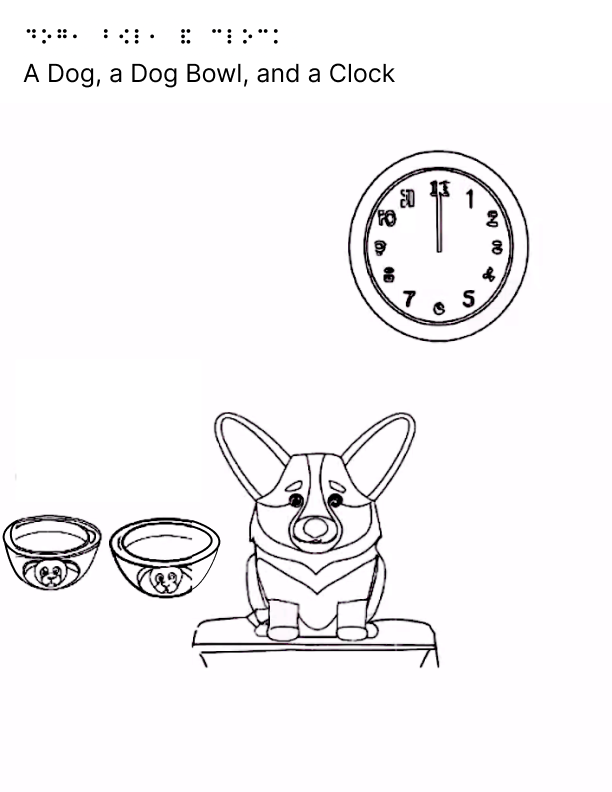}
\Description{The image presents three sample user graphic generations for 'Task One'. Each sample depicts three items: a dog, a dog bowl, and a clock. The first graphic shows a simple, cartoon-style representation with a puppy in front of a dog bowl that has a few pieces of dog food scattered around, and a wall clock with roman numerals above. In the second graphic, there is a more detailed drawing of a dachshund with a wagging tail next to an empty dog bowl, and a wall clock with hands pointing at eleven and one.
The third graphic features a corgi sitting proudly between two dog bowls decorated with paw prints, and a wall clock with a second hand above it.}

\subsection{TaskTwo: Sample User Graphic Generations}
\label{sec:user_graphic}
\includegraphics[width=0.33\textwidth]{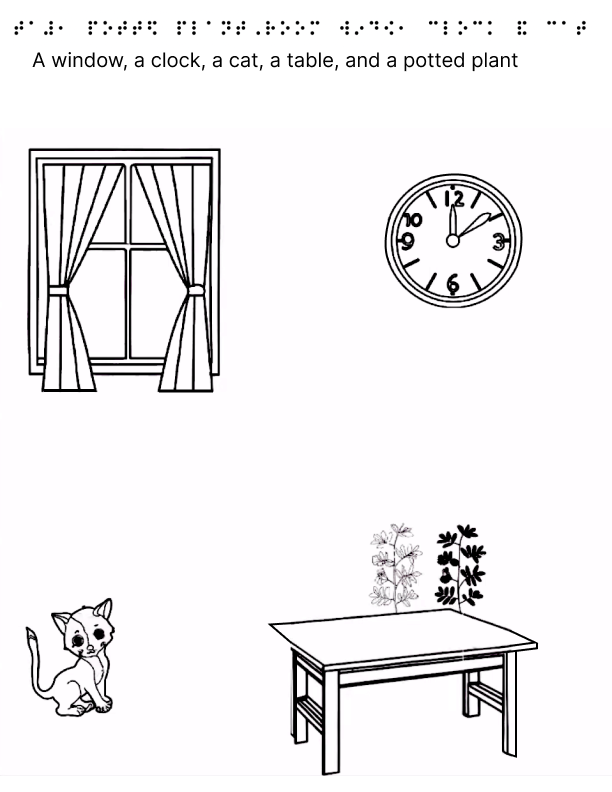}
\includegraphics[width=0.33\textwidth]{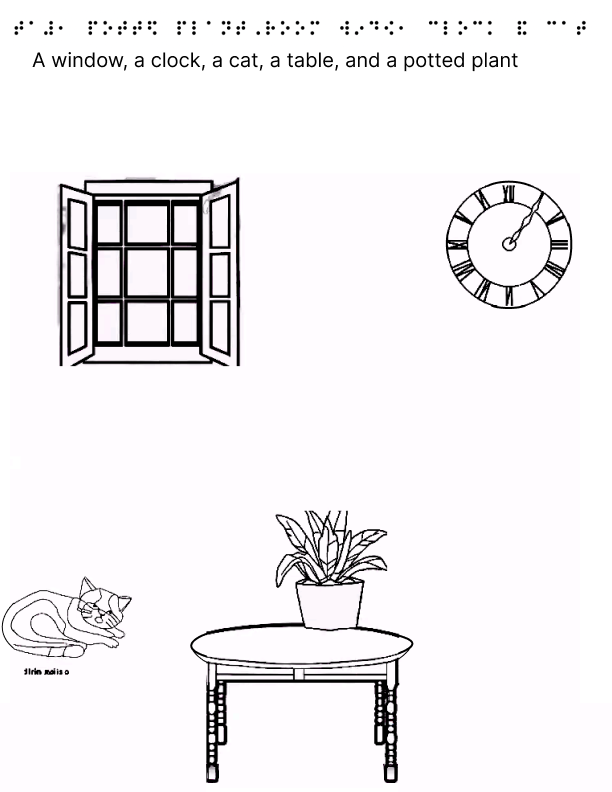}
\includegraphics[width=0.33\textwidth]{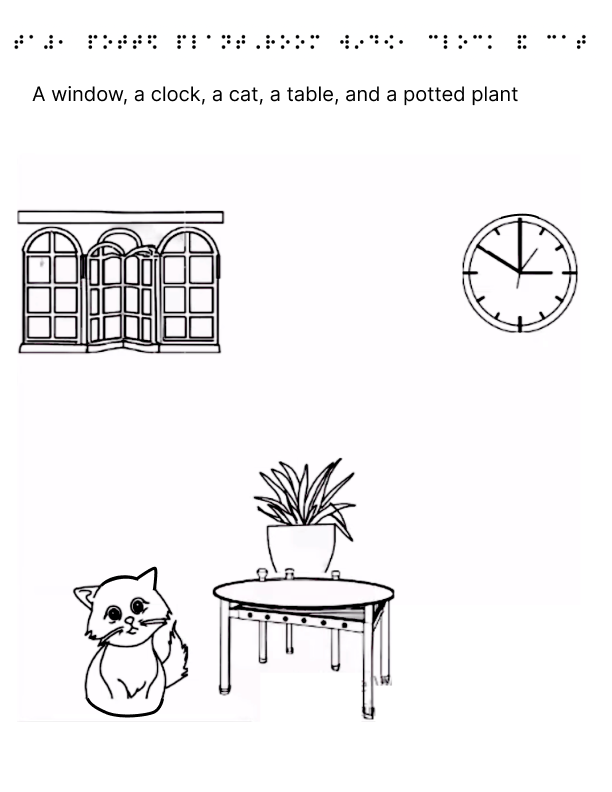}
\Description{ The image showcases three sets of line-drawn graphics created by users as part of 'Task Two'. Each set includes a window, a clock, a cat, a table, and a potted plant. In the first graphic, the window is styled with curtains pulled aside, the clock has simple numbers and hands, the cat appears seated and alert, the table is square with a flowerpot that has several flowers.
The second graphic features a grid-style window, a clock with Roman numerals, a lounging cat, and a round table with a large potted plant.
The third graphic depicts an arched window with a divided pane, a clock marked with lines instead of numbers, a fluffy cat seated in front of a slender table supporting a potted plant with long leaves.
Each graphic presents these common elements with varying artistic interpretations, showcasing different styles and details in the users' renditions.}

\subsection{TaskThree: Sample User Graphic Generations}
\label{sec:user_graphic}
\includegraphics[width=0.33\textwidth]{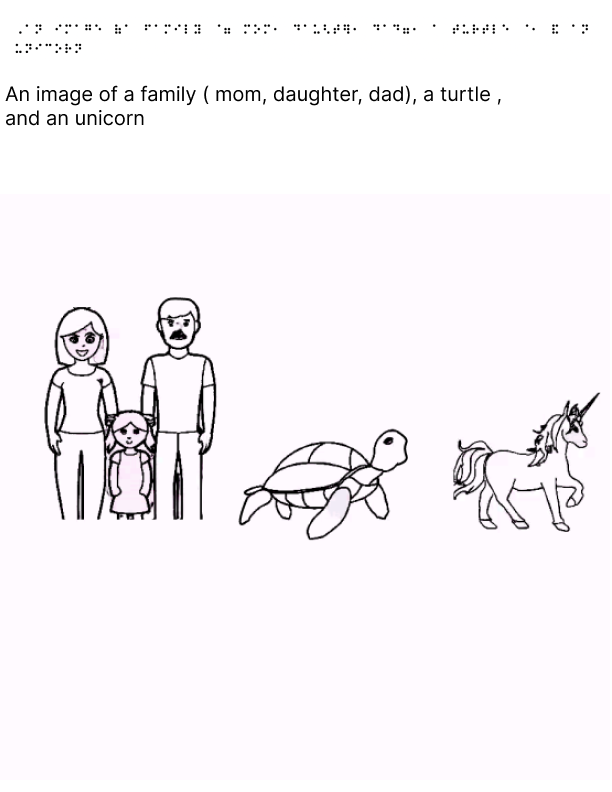}
\includegraphics[width=0.33\textwidth]{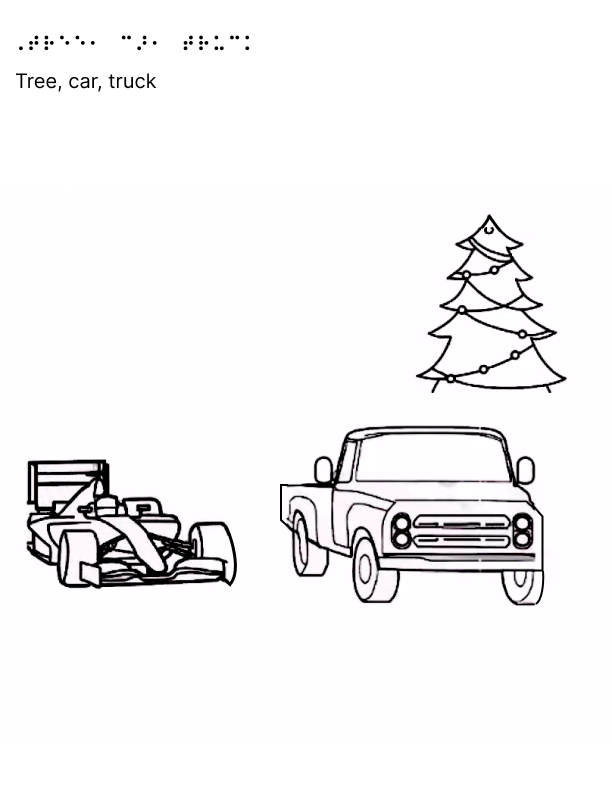}
\includegraphics[width=0.33\textwidth]{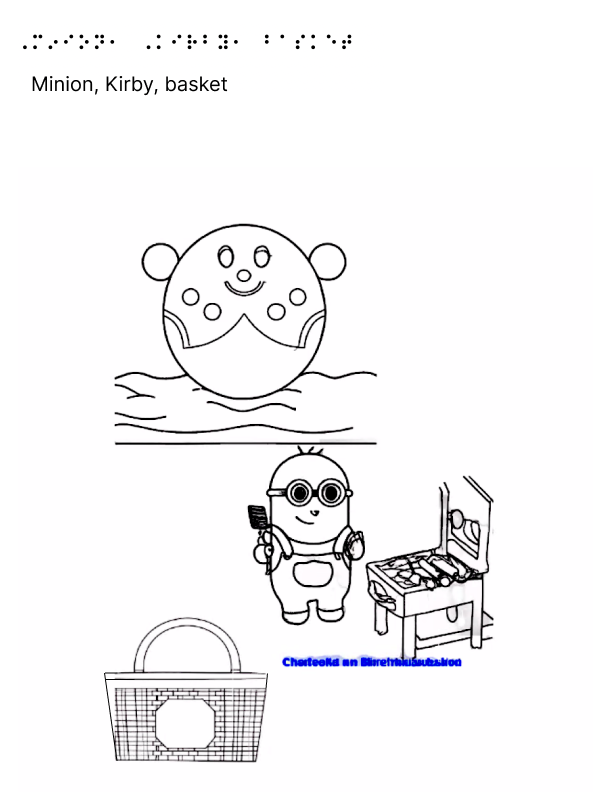}
\Description{
The image is a compilation of three sets of graphics generated by users for 'Task Three'. Each graphic represents a different theme described by the users: The first graphic depicts a family of three (mom, dad, and daughter), standing next to a turtle and a unicorn, showcasing a combination of real and mythical creatures.
The second graphic shows a scene with three items: a decorated Christmas tree, a racing car, and a pickup truck, suggesting a contrast between the stationary and the dynamic.
The third graphic features a playful set with a character that resembles a Minion wearing glasses and holding a spatula grilling things for a barbecue, a round character that could be Kirby floating on water, and a simple basket, highlighting characters from popular culture and a common object.
Each set of images captures unique elements as specified by the users, reflecting their individual creative interpretations of the given task.}

\subsection{Prompts used in the system}
\label{sec:prompts}

\subsubsection{Tactile Image Generation Prompt}

\begin{promptbox}
\begin{verbatim}
Create ONLY ONE DIGITAL graphic of the \texttt{${mainObject}} 
This graphic should be VERY SIMPLE and MINIMAL, focusing on the core shape and essence of the object. 
Avoid adding any perspectives, intricate details, or text to the design. 
The goal is to capture the simplicity and clarity of the object in a minimalistic style.
Ensure that the lines are clean and the overall design is straightforward. 
The user has specifically requested the following object: \texttt{${voiceText}}. 
Please adhere strictly to these guidelines to achieve the desired outcome.
\end{verbatim}
\end{promptbox}

\subsubsection{Image Description Prompt}

\begin{promptbox}
\textbf{GLOBAL DESCRIPTION:}
\begin{verbatim}
Provide a one-line brief description of what the image looks like. 
Describe the layout of the images on a canvas based on their coordinates
and sizes verbally, without using exact numbers. Avoid stating specific
shapes like "square." Mention if one image appears to be placed on top
of another, noting any spaces above or below the image.
\end{verbatim}
\textbf{LOCAL DESCRIPTION:}
\begin{verbatim}
The image is called \texttt{\$\{image.name\}}. It is located at x-coordinate
\texttt{\$\{image.coordinate.x\}} and y-coordinate \texttt{\$\{image.coordinate.y\}}.
The size of the image is \texttt{\$\{image.sizeParts.width\}} in width
and \texttt{\$\{image.sizeParts.height\}} in height. 
Additional description: \texttt{\$\{image.descriptions\}}.    
\end{verbatim}
\textbf{CHAT BASED DESCRIPTION:}
\begin{verbatim}
You are describing an image to a Visually Impaired Person. Keep the description brief
and straightforward. Generate the given image description according to the
following criteria: \texttt{\$\{voiceText\}}.
\end{verbatim}
\end{promptbox}

\end{document}